\documentclass[journal,twoside,web]{ieeecolorArxiv}
\usepackage{tmi_nologo}
\usepackage{url}
\usepackage{tabularx}
\usepackage{amsmath,amssymb,amsfonts}
\usepackage{cite}
\usepackage[font=small,labelfont=bf]{caption}
\usepackage{subcaption}
\usepackage{siunitx}
\usepackage[artemisia]{textgreek}
\usepackage{newunicodechar}

\usepackage[utf8]{inputenc}
\usepackage{color}
\usepackage{multirow}
\usepackage{chngcntr}

\usepackage{algorithmic}
\usepackage{graphicx}
\usepackage{textcomp}

\usepackage{xr-hyper}

\definecolor{magenta}{RGB}{139,0,139}
\def\baselinestretch{1.00}
\def\BibTeX{{\rm B\kern-.05em{\sc i\kern-.025em b}\kern-.08em
    T\kern-.1667em\lower.7ex\hbox{E}\kern-.125emX}}
{
}
\begin{document}
\title{TranSMS: Transformers for Super-Resolution Calibration in Magnetic Particle Imaging}  \vspace{-0.75cm}
\author{Alper G\"{u}ng\"{o}r, Baris Askin, Damla Alptekin Soydan, Emine Ulku Saritas, Can Bar\i \c{s} Top,\\ Tolga \c{C}ukur, \IEEEmembership{Senior Member} \vspace{-1.00cm} 
\\
\thanks{This study was supported in part by TUBA GEBIP 2015 and BAGEP 2017 fellowships. The work of Alper G\"{u}ng\"{o}r was supported by the Scientific and Technological Research Council of Turkey (TÜBİTAK) under Grant BİDEB 2211. Corresponding author: A. G\"{u}ng\"{o}r (alperg@ee.bilkent.edu.tr).}
\thanks{A. G\"{u}ng\"{o}r, B. Askin, E.U. Saritas, and T. \c{C}ukur are with the Department of Electrical and Electronics Engineering, and National Magnetic Resonance Research Center, Bilkent University, Ankara, Turkey (e-mails: \{alperg, baris.askin, saritas, cukur\}@ee.bilkent.edu.tr). A. G\"{u}ng\"{o}r is also with Aselsan Research Center, Ankara, Turkey. D.A. Soydan, and C.B. Top are with Aselsan Research Center, Ankara, Turkey (e-mails: \{dasoydan, cbtop\}@aselsan.com.tr).}
}

\maketitle

\begin{abstract}
Magnetic particle imaging (MPI) offers exceptional contrast for magnetic nanoparticles (MNP) at high spatio-temporal resolution. A common procedure in MPI starts with a calibration scan to measure the system matrix (SM), which is then used to set up an inverse problem to reconstruct images of the MNP distribution during subsequent scans. This calibration enables the reconstruction to sensitively account for various system imperfections. Yet time-consuming SM measurements have to be repeated under notable changes in system properties. Here, we introduce a novel deep learning approach for accelerated MPI calibration based on Transformers for SM super-resolution (TranSMS). Low-resolution SM measurements are performed using large MNP samples for improved signal-to-noise ratio efficiency, and the high-resolution SM is super-resolved via model-based deep learning. TranSMS leverages a vision transformer module to capture contextual relationships in low-resolution input images, a dense convolutional module for localizing high-resolution image features, and a data-consistency module to ensure measurement fidelity. Demonstrations on simulated and experimental data indicate that TranSMS significantly improves SM recovery and MPI reconstruction for up to 64-fold acceleration in two-dimensional imaging.
\end{abstract}


\bstctlcite{IEEEexample:BSTcontrol}

\section{Introduction}
\label{sec:introduction}

Magnetic particle imaging (MPI) is a recent imaging modality to map \textit{in vivo} distribution of magnetic nanoparticles (MNPs) with exceptional sensitivity and speed \cite{gleich_tomographic_2005,saritas2012rev}. Preclinical studies have readily demonstrated its potential in angiography, stem cell tracking, cancer imaging, neuroimaging, and localized hyperthermia  \cite{Weizenecker_2009,zheng2015magnetic,arami2017tomographic,song2017janus,utkur2017,ludewig2017magnetic,cooley2018rodent,tay2018magnetic,tong2021highly}. MPI acquires the nonlinear magnetization response of MNPs to an oscillating drive field (DF) while a static selection field (SF) creates a field-free region (FFR) for spatial encoding. Time-domain signals are recorded on receive coils while the FFR is traversed across a field-of-view (FOV). The forward mapping from MNP concentration to recorded signals depends on a system function that captures properties regarding MNP dynamics, scan trajectories, and scanner hardware. Therefore, MPI reconstruction relies on estimates of this system function to infer the MNP distribution from time-domain signals.

Two fundamental approaches for MPI reconstruction differ in their treatment of the system function: X-space and system matrix (SM) methods. \textcolor{black}{In X-space methods, a point spread function is used to characterize the MPI signal, typically based on an analytical model or alternatively based on measurements \cite{GoodwillXspace2011}. In contrast, while analytical approaches have been proposed \cite{superResolutionMPISM}, SM methods commonly measure the system function using a calibration scan where a small MNP sample spanning an imaging voxel is traversed on a spatial grid.} When the system function is directly measured on the imaging system, SM methods offer improved immunity against system imperfections \cite{ilbey2017comparison, gruttner2013formulation}. Yet, a time-consuming calibration (e.g., $\sim$12 hours for a typical 32$\times$32$\times$32 grid) must be repeated whenever the MNP type or scanning trajectory is changed \cite{von2017hybrid}. Pioneering studies in this domain have proposed accelerating calibration based on compressed sensing (CS) \cite{Lampe_2012, Knopp_2013, Weber_2015, Kaethner_nonequal_calibration, FastCalibrationSilbey} and super-resolution (SR) \cite{superResolutionMPISM, 2d-SMRnet, omer_2015}. However, calibration using a small MNP sample can manifest low signal-to-noise ratio (SNR) efficiency, limiting the quality of recovered SMs at high acceleration rates \cite{FastCalibrationSilbey, Gungor_iwmpi}.

Here we propose a novel deep learning (DL) method for accelerated MPI calibration, named Transformers for SM Super-resolution (TranSMS). For improved SNR efficiency, TranSMS performs low-resolution (LR) SM measurements using a larger MNP sample spanning across an LR voxel as opposed to a high-resolution (HR) voxel. HR SM is then recovered by a deep network that super-resolves LR SM. \textcolor{black}{SM rows follow spatial patterns that visually resemble sinusoidal grating functions, windowed by an envelope whose shape is linearly proportional to the receive coil sensitivity. Thus, pixel intensities in each SM intrinsically show long-range correlations.} \textcolor{black}{Prior SR methods are predominantly based on convolutional neural networks (CNNs) that improve performance over traditional methods \cite{2d-SMRnet,embc}. However, due to their local kernels, CNNs cannot capture the long-range contextual features in SM rows. To address this limitation,} we introduce a novel architecture that aggregates the localization power of CNNs with the contextual expressiveness of vision transformers. TranSMS leverages a novel data-consistency module to ensure fidelity to LR SM measurements based on the signal model between LR and HR SM. Comprehensive demonstrations are performed on simulated and experimental data from field-free point (FFP) and field-free line (FFL) scanners, for up to 64-fold acceleration during two-dimensional (2D) imaging. Our results indicate that TranSMS offers substantial performance improvements in both SM recovery and image reconstruction over the state-of-the-art CS- and SR-based calibration methods. Our main contributions are:
\begin{itemize}
    \item We introduce a novel deep learning approach to super-resolve MPI SMs for accelerated calibration.
    \item To our knowledge, we introduce the first vision transformer model for MPI. 
    \item We introduce a novel data-consistency module to incorporate the physical signal model relating HR SM to LR SM measurements. 
\end{itemize}

\section{Related Work}
Several successful approaches have previously been introduced for accelerating MPI calibration. A hybrid approach combines model- and measurement-based SM estimates \cite{superResolutionMPISM}. An HR SM is analytically simulated while an LR SM is measured. This method was used to achieve 4-fold accelerated 2D calibration by 2-fold increase of voxel dimensions in LR SM. Reconstruction is then formulated as a weighted optimization with terms reflecting the simulated and measured SMs. This model-assisted SR approach derives HR information from an analytical model. Thus, accuracy might be compromised by system imperfections that are not captured by the model.

The CS framework instead exploits sparsity to recover signals from undersampled measurements \cite{Lampe_2012, Knopp_2013, Weber_2015}. Each SM row is a complex-valued spatial sensitivity map with compressible representations in Fourier, cosine or Chebyshev polynomial domains \cite{Lampe_2012}. Thus, calibrations can be accelerated by randomly undersampling the grid locations during HR SM measurements (Fig.~\ref{fig:imagingOut}b). The full SM can then be recovered by simultaneously enforcing consistency to available measurements, transform-domain sparsity, and additional priors regarding the symmetry of sensitivity maps. For 3D FFP imaging, accelerations up to 27-fold were demonstrated under relatively limited noise \cite{Weber_2015}. Other optimization-based methods have exploited non-equispaced sampling and low-rank approximations through higher-order singular value decomposition \cite{Kaethner_nonequal_calibration, hosvd_calibration}. Yet, HR SM measurements using small MNP samples spanning a single imaging voxel limit SNR efficiency. Although calibration with multiple MNP samples per measurement can improve efficiency \cite{FastCalibrationSilbey}, there are practical challenges in positioning of a large calibration phantom.

Recently, DL techniques have also been considered for MPI calibration \cite{2d-SMRnet}. A previous study has proposed using strided LR SM measurements, as shown in Fig.~\ref{fig:imagingOut}c, followed by upsampling to recover an HR SM \cite{2d-SMRnet}. Upsampling is performed via a CNN trained using paired sets of LR and HR SMs. This SR approach was employed to recover HR SM from 64-fold accelerated 3D calibration scans (analogously 16-fold acceleration for 2D calibration). That said, strided LR SM measurements performed using an MNP sample spanning a single HR voxel can potentially limit SNR efficiency. 

\textcolor{black}{Super-resolving HR SM in MPI is closely related to the single-image super-resolution (SISR) problem in computer vision \cite{superResolutionMPISM, 2d-SMRnet, embc}. DL methods have recently become the gold standard in this domain. Successful SISR models have been introduced for natural images based on convolutional architectures, such as Super-Resolution Convolutional Neural Network (SRCNN) \cite{srcnn}, Very Deep Super-Resolution (VDSR) \cite{vdsr}, 
and Residual-Dense Network (RDN) \cite{rdn}. 
That said, characteristics of SM rows notably differ from natural images. First, SM sizes are relatively compact compared to natural images. Second, SMs contain a single broadly distributed pattern whereas multiple nonoverlapping objects can exist in natural images. Third, SMs are corrupted by higher levels of noise compared to natural images. Due to these differences, deep architectures with high-dimensional feature maps can be suboptimal for SM recovery.} Furthermore, CNN models with compact filters have limited sensitivity to long-range contextual features \cite{reviewPaperMedicalSR}. \textcolor{black}{This has motivated the adoption of hybrid CNN-transformer models for improved contextual sensitivity in SISR \cite{hybridmodel1,hybridmodel2}. However, prior models used serially-connected CNNs and vanilla vision transformers \cite{ViT}. In contrast, TranSMS uses parallel-connected convolutional and transformer modules, and novel transformer blocks with convolutional patch embedding. Few recent studies have introduced efficient transformers for medical imaging tasks \cite{transUnet,karimi2021,korkmaz2021unsupervised,dalmaz2021resvit}. Yet, this is the first study to introduce a vision transformer for MPI, and the first study to introduce a model-based transformer for SR.}

\section{Theory}
\subsection{Magnetic Particle Imaging}

\subsubsection{Signal Model}
In MPI, excitation is performed via an oscillating DF at a fundamental frequency, while spatial encoding is enabled via a static SF that creates an FFR within the FOV. Voltage waveforms captured by receive coils reflect MNP responses within the FFR. SM-based reconstruction is performed by transforming the received time domain signal to frequency domain. The signal model can be expressed as \cite{gleich_tomographic_2005}:   
\begin{align}
    \mathbf{Ax+n = y} \label{eq:mpiModel}
\end{align}
where $\mathbf{A} \in C^{M \times N}$ is the SM; $\mathbf{x} \in \mathbb{R}^{N}$, $\mathbf{n} \in C^{M}$ and $\mathbf{y} \in C^{M}$ are the image vector, noise vector and frequency domain signal, respectively. Here, $M$ is the number of frequency components and $N$ is the number of voxels in the grid. MPI data contain thermal noise that varies as a function of frequency \cite{Paysen_2020}, and the overall signal intensity decreases with increasing \textcolor{black}{harmonic} frequency \cite{gleich_tomographic_2005}. Thus, the additive noise in Eq.~\eqref{eq:mpiModel} can be taken as colored Gaussian \cite{Paysen_2020, cbtop2020}.

\subsubsection{SM Calibration}
\begin{figure}[t]
\centering
	\includegraphics[width=0.8\columnwidth]{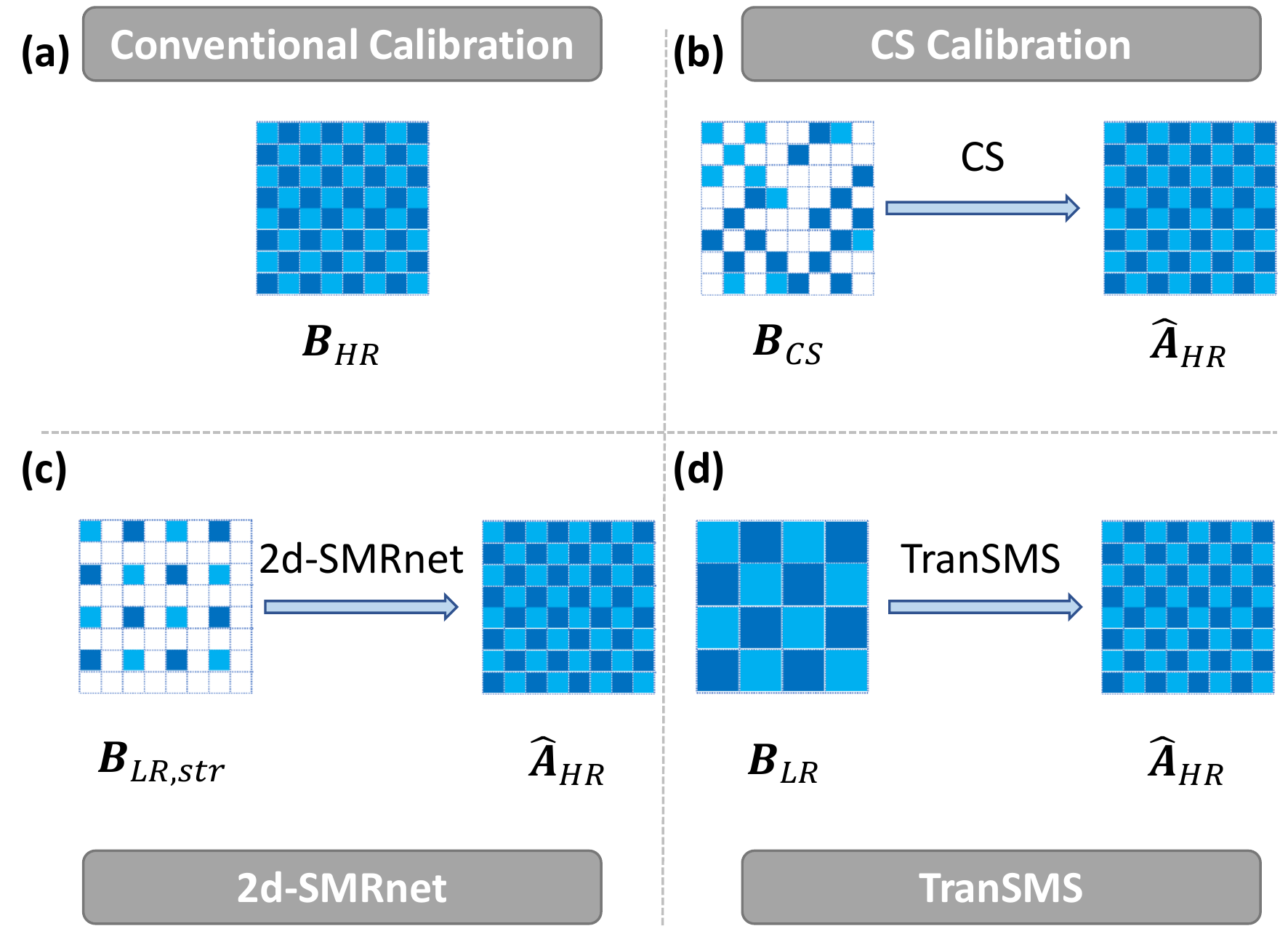}
\caption{Overview of prior and proposed techniques for SM calibration. \textbf{(a)} Conventional calibration samples all of the grid points in HR SM. \textbf{(b)} CS calibration randomly subsamples HR SM and uses CS optimization for recovery. \textbf{(c)} 2d-SMRnet uniformly subsamples HR SM in a strided fashion and recovers HR SM via a CNN. \textbf{(d)} TranSMS performs LR SM measurements with a larger MNP sample that fills an LR grid, and recovers HR SM via a transformer.
}
	\label{fig:imagingOut}
\end{figure}

The SM  is affected by imperfections such as system drift or inter-batch variability in MNP response \cite{knopp2012magnetic}, and is spatially variant \cite{Knopp_spatially_variant}. To account for such non-idealities, SM calibration is performed by traversing a small MNP sample across the imaging grid \cite{von2017hybrid}, as shown in Fig.~\ref{fig:imagingOut}a:
\begin{align}
    \mathbf{B} = \mathbf{A + \Tilde{N}}. \label{eq:CSmodel}
\end{align}
Here, $\mathbf{\Tilde{N}}$ and $\mathbf{B}$ denote additive noise and the measured SM, respectively. Measurements are typically taken with an MNP sample size that matches the intended imaging voxel size. This limits SNR efficiency in HR SM measurements.

\subsubsection{Image Reconstruction}
Once an SM estimate $\mathbf{\hat{A}}$ \textcolor{black}{is obtained from measurements $\mathbf{B}$}, image reconstruction can be cast as an inverse problem to solve Eq.~\eqref{eq:mpiModel}. Two common optimization approaches are the Kaczmarz \cite{Kaczmarz} and alternating direction method of multipliers (ADMM) methods \cite{ilbey2017comparison, FastCalibrationSilbey}. Here, we consider the ADMM-based formulation \cite{simitcs, boydADMM}:
\begin{align}
    \arg\min_{\mathbf{x}} \alpha_1 \Vert \mathbf{x} \Vert_1 + \alpha_{TV} \text{TV}(\mathbf{x}) \text{ s.t. } \Vert \mathbf{\hat{A}x - y} \Vert_2\leq \epsilon, \label{eq:imageRecon}
\end{align}
where $\alpha_{1,TV}$ are regularization weights. Eq.~\eqref{eq:imageRecon} promotes sparsity in image and finite-differences (TV) domains, while constraining fidelity of reconstructed image and acquired signals with an $\epsilon$ bound based on noise level. To account for varying noise levels across SM rows in Eq.~\eqref{eq:imageRecon}, repeated background measurements can be used for noise whitening, and filtering out frequency components with low SNR \cite{Knopp_2015_background, Knopp_2021_background, otherReconMPI, cbtop2020}. \textcolor{black}{For whitening, $\mathbf{\hat{A}}$ and $\mathbf{y}$ can be multiplied by a diagonal matrix with elements $\mathbf{E}_{i,i} = 1/\sigma_i$, where $\sigma_i$ is the standard deviation (std.) of noise for the $i^{th}$ SM row (i.e., std. of noise in $i^{th}$ component in frequency domain)} \cite{otherReconMPI, cbtop2020}. Here, we pre-whiten all SM rows to unit std.

\begin{figure*}[ht]
	\centering
	\includegraphics[width=0.95\textwidth]{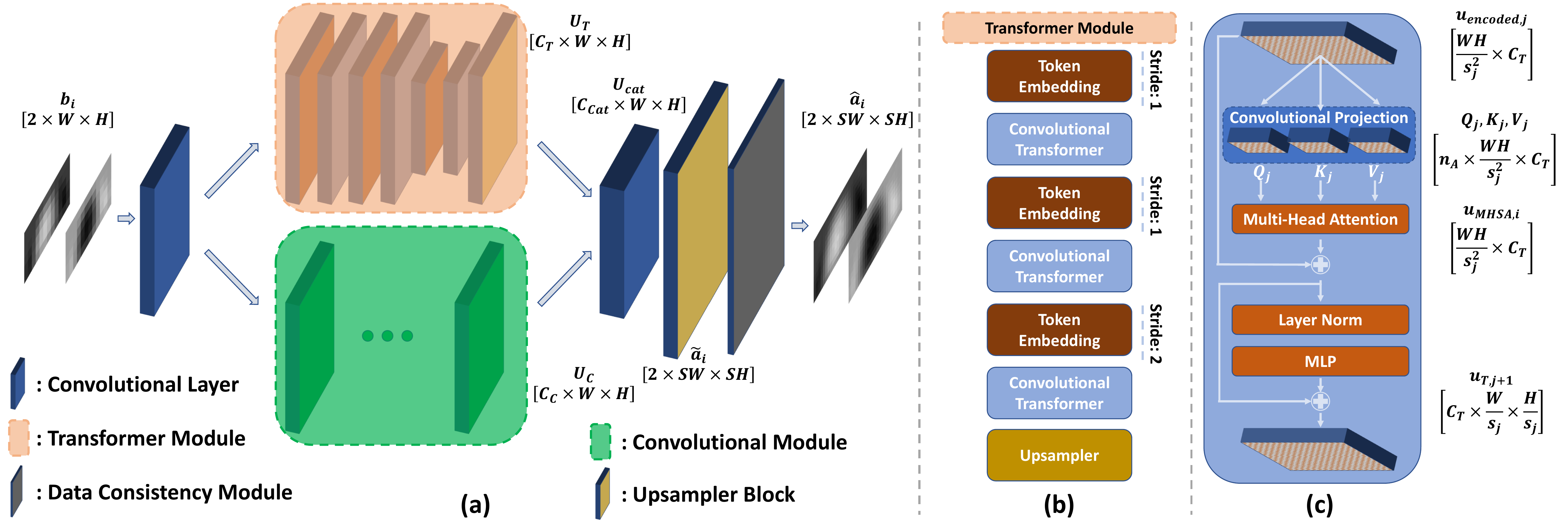}
	\caption{
	\textbf{(a)} Schematic of \textcolor{black}{the network architecture in} TranSMS. After an initial convolutional layer, encoded feature maps are fed to parallel transformer and convolutional modules. The transformer module extracts contextual representations over longer distances, while the convolutional module captures localized representations. Contextual and localized feature maps are fused and upsampled to the resolution of the HR SM. Finally, a data-consistency module enforces fidelity of recovered HR SM to input LR SM. \textbf{(b)} The transformer module contains a cascade of token-embedding and convolutional transformer sub-blocks, and a final upsampler to restore dimensionality. \textbf{(c)} The transformer performs convolutional token projection, followed by MHSA and MLP blocks to extract contextual features.}
	\label{fig:TranSMS_general_structure}
\end{figure*}

\subsection{TranSMS}
\textcolor{black}{To improve SNR efficiency, TranSMS performs LR SM measurements using larger MNP samples as opposed to smaller samples used in strided LR SM measurements (Fig.~\ref{fig:imagingOut}). To improve HR SM recovery, the SR problem is solved via a novel transformer architecture.} Code for TranSMS will be available at {\small \url{https://github.com/icon-lab/TranSMS}}.

\subsubsection{Super-Resolution MPI Calibration}
Inspired by recent reports \cite{superResolutionMPISM, Gungor_iwmpi}, here we propose to measure LR SM with MNP samples matching the LR SM voxel size. A large MNP sample spans multiple voxels on the HR SM grid, so the proposed measurements correspond to box-car downsampling of the HR SM with a matrix $\mathbf{D}$:
\begin{align}
    \mathbf{B} = \mathbf{A D^T + \Tilde{N}} \label{eq:superResolvingModel}
\end{align}
For each SM row, the sampling relationship is $   \mathbf{b}_i = \mathbf{D} \mathbf{a}_i + \mathbf{n}_i,$ so recovery of HR SM row $\mathbf{a}_i = (\mathbf{A}_{HR})_{i}$ given LR SM row $\mathbf{b}_i$ is an SR task. \textcolor{black}{Since noise does not depend on the MNP sample size, the proposed measurements yield improved SNR efficiency.} For instance, measurements with a 16-fold larger MNP sample would result in a 16-fold SNR increase.

\subsubsection{Network Model}
We solve Eq.~\eqref{eq:superResolvingModel} to recover HR SM from LR SM measurements. MPI SMs resemble sinusoidal gratings with strong inter-voxel correlations over broad distances. The MPI signal also shows spatially-variant characteristics due to coil inhomogeneities. Taken together, these factors render CNN-based SR with local, translation-invariant filters suboptimal for SM recovery. \textcolor{black}{Instead, we propose a hybrid architecture to super-resolve HR SM at an SR factor of $S$, corresponding to $S^2$-fold acceleration for 2D calibration}, leveraging transformer, convolutional, and data-consistency modules. The transformer module captures contextual image features, the convolutional module captures localized features, whereas the data-consistency module enforces fidelity to LR SM measurements. The network input is a reformatted row of LR SM $\mathbf{b}_i \in \mathbb{R}^{2 \times W \times H}$, with two channels in the first dimension representing real/imaginary parts, $W$ and $H$ denoting the width and height of the LR grid (i.e., $W H = N/S^2$). An initial convolutional layer $Z_{init}(\cdot)$ expands the channels to produce $\mathbf{U}_{init}  = Z_{init}(\mathbf{b}_i) \in \mathbb{R}^{C_1 \times W \times H}$ (with $C_1$ channels), which is then fed to the transformer and convolutional modules.

\textbf{Transformer module:} 
The transformer module contains a cascade of $n_{TB}$ submodules that capture contextual features in SM rows (Fig.~\ref{fig:TranSMS_general_structure}). Each submodule has a token-embedding block ($B_{TE,j}(\cdot)$) followed by a transformer block ($B_{T,j}(\cdot)$), where $j$ denotes submodule index. Token-embedding is achieved via a sequence of convolutional, reshaping and layer normalization (LN) layers \cite{CvT}. Assuming the input is $\mathbf{U}_{T,j-1}$ (where $\mathbf{U}_{T,0} = \mathbf{U}_{init}$), the token-embedded feature map is: 
\begin{align}
    \mathbf{u}_{\text{encoded},j} &= \text{LN}(\text{Reshape}_\text{1D}(\text{Conv}_\text{2D}(\mathbf{U}_{T,j-1}))), \label{eq:embedding}
\end{align}
where $\text{Conv}_\text{2D}$ has $3\times3$ kernels with stride $s_j$. Unlike non-overlapping patch embeddings \cite{ViT}, overlapping token embeddings in Eq.~\eqref{eq:embedding} can improve the capture of local and global context \cite{CvT}. $\mathbf{u}_{\text{encoded},j} \in \mathbb{R}^{WH / s_j^2 \times C_T}$ ($C_T$ denoting number of channels) is processed via a \textit{convolutional transformer} block comprising convolutional projection, multi-head self-attention (MHSA) and multi-layer perceptron (MLP) sub-blocks. The projection extracts query, key and value for each token via depth-wise separable convolution (DWSC) \cite{DWSC}: 
\begin{align}
    \mathbf{Q}_j, \mathbf{K}_j, \mathbf{V}_j &= \text{Flatten}(\text{DWSC}(\text{Reshape}_\text{2D}(\mathbf{u}_{\text{encoded}, j})),
\end{align}
where the reshape operator formats the dimensions to $\mathbb{R}^{C_T \times W/s_j \times H / s_j}$, $\mathbf{Q}_j, \mathbf{K}_j, \mathbf{V}_j \in \mathbb{R}^{n_A \times WH/ s_j^2 \times  C_T }$ (with $n_A$ denoting the number of attention heads) are query, key and value, respectively. $\text{DWSC}$ is implemented as:
\begin{align}
   \text{DWSC}\left( \mathbf{v}\right) &= \text{Conv}_\text{1}(\text{BN}(\text{Conv}_\text{2D}(\mathbf{v}) )),
\end{align}
where $\text{Conv}_\text{1}(\cdot)$ is $1 \times 1$ convolution, BN is batch normalization. The projected maps are then residually fed to MHSA:
\begin{align}
    \mathbf{u}_\text{MHSA,j} &= \text{MHSA}(\mathbf{Q}_j, \mathbf{K}_j, \mathbf{V}_j) + \mathbf{u}_{\text{encoded},j}, 
\end{align}
where $\mathbf{u}_\text{MHSA,j} \in \mathbb{R}^{WH / s_j^2 \times C_T}$ is the output and MHSA is:
\begin{align}
    \text{MHSA} &= \text{Lin} \left( \text{Reshape}_\text{1D} \left( \text{SftMx}\left( \mathbf{Q}_j \mathbf{K}_j^T \right) \mathbf{V}_j \right) \right), \label{eq:mhsa}
\end{align}
In Eq.~\eqref{eq:mhsa}, the linear layer (Lin) compresses the number of channels from $n_A C_T$ to $C_T$. Finally, the feature maps are projected with residual connection through an MLP:
\begin{align}
    \mathbf{u}_{T,j+1} &= \text{Reshape}_{\text{2D}}(\text{MLP}(\text{LN}(\mathbf{u}_\text{MHSA,j})) + \mathbf{u}_\text{MHSA,j}), 
\end{align}
where $\mathbf{u}_{T,j+1} \in \mathbb{R}^{C_T \times W / s_j \times H / s_j}$ is the output map, and MLP has a single hidden layer of size $2C_T$. A stride of $s_j>1$ broadens the convolutional kernels to permit analysis at a larger scale. Here, we used $n_{TB} = 3$ submodules with $s_j = [1, 1, 2]$. Since the last submodule halved the image size, an upsampler block (see Fusion Module for details) was used to restore the original dimensions of $W \times H$.

\textbf{Convolutional module:} The convolutional module is composed of convolutional layers ($Z_{C,0}(\cdot)$ \& $Z_{C,1}(\cdot)$) followed by a cascade of $n_{RDB}$ Residual-Dense Blocks (RDBs). Convolution layers project $\mathbf{U}_{init}$ to feature map $\mathbf{u}_0 \in \mathbb{R}^{C_C \times W \times H}$.
\begin{align}
    \mathbf{u}_{-1} = Z_{C,0}(\mathbf{U}_{init}), ~ \mathbf{u}_0 = Z_{C,1}(\mathbf{u}_{-1})
\end{align}
Then, the RDBs sequentially process the feature maps as: 
\begin{align}
    \mathbf{u}_d &= B_{d}(\mathbf{u}_{d-1})
\end{align}
where $B_{d}(\cdot)$ is the projection through the $d^{th}$ RDB with $n_{CL}$ convolutional layers each performing a projection  $Z_{d,k}(\cdot)$. Each layer concatenates all feature maps from previous layers:
\begin{align}
    \mathbf{u}_{d,k} &= Z_{d,k}([\mathbf{u}_{d-1}; \mathbf{u}_{d,1}; \mathbf{u}_{d,2}; \cdots; \mathbf{u}_{d,k-1}])
\end{align}
where $\mathbf{u}_{d,k}$ has $n_{GR}$ channels denoting the growth-rate of the block across a layer. The last layer implements $Z_{d,out}(\cdot)$ to reduce the number of channels to $C_C$ via a $1 \times 1$ convolution:
\begin{align}
    \mathbf{u}_d &= Z_{d,out}([\mathbf{u}_{d-1}; \mathbf{u}_{d,1}; \cdots; \mathbf{u}_{d,r}]) + \mathbf{u}_{d-1}
\end{align}
%
%
Finally, the output of all RDBs are concatenated through a $1 \times 1$ convolution operator $Z_{C,ct}$ as:
\begin{align}
    \mathbf{u}_{ct} &= Z_{C,ct}([\mathbf{u}_{1}; \mathbf{u}_{2}; \cdots; \mathbf{u}_{n_{RDB}}])
\end{align}
The output is then combined with a global residual from the initial layer using a final convolutional layer $Z_{C,last}(\cdot)$ as:
\begin{align}
    \mathbf{U}_C &= Z_{C,last}(\mathbf{u}_{ct}) + \mathbf{u}_{-1}
\end{align}
where $\mathbf{U}_C$ denotes the output of the convolutional module.

\textbf{Fusion module:} 
The fusion module first aggregates localized and contextual representations, $\mathbf{U}_C$ and $\mathbf{U}_T$, respectively from the convolutional and transformer modules via a convolutional layer ($Z_{cat}(\cdot)$) to generate $\mathbf{U}_{cat} \in \mathbb{R}^{C_{cat} \times H \times W}$:
\begin{align}
    \mathbf{U}_{cat} &= Z_{cat}([\mathbf{U}_{C}; \mathbf{U}_{T}])
\end{align}
The fused map is then processed via a cascade of 2-fold upsampling blocks. Assuming that S is a power of 2, a feature map $\mathbf{U}_{ups} \in \mathbb{R}^{C_{cat} \times SH \times SW}$ is generated:
\begin{align}
    \mathbf{U}_{ups} &= \text{Upsampler}_{2{\times}}(\cdots(\text{Upsampler}_{2{\times}}(\mathbf{U}_{cat})))
\end{align}
The upsampling blocks consist of convolutional layers that expand the number of channels 4-fold, and a pixel shuffler to map channels onto higher resolution tensors \cite{espcn}. 
A final convolutional layer ($Z_{fin}(\cdot)$) produces $\Tilde{\mathbf{a}}_i \in \mathbb{R}^{2 \times S H \times S W}$, the $i^{th}$ row of SM super-resolved by a factor of $S$:
\begin{align}
    \Tilde{\mathbf{a}}_i &= Z_{fin}(\mathbf{U}_{ups})
\end{align}

\textbf{Data-Consistency Module:}
DL methods for SR tasks typically perform a direct mapping between LR and HR images without observing a signal model\textcolor{black}{, posing a risk for inconsistency between measured LR and predicted HR images. Although the training loss can implicitly satisfy consistency on the training set, it offers no guarantees on the test set. To obtain consistent LR-HR SMs,} we introduce a novel data-consistency (DC) module in TranSMS to enforce fidelity between recovered and measured data as inspired by recent MRI methods \cite{salman_2020, Dong_liang}. Based on the signal model in Eq.~\eqref{eq:superResolvingModel}, DC can be enforced by comparing box-car downsampled HR SM with measured LR SM. Receiving as input $\Tilde{\mathbf{a}}_i$:
\begin{align}
    \hat{\mathbf{a}}_{i} &= \arg \min_{\mathbf{a}} \Vert \mathbf{a} - \Tilde{\mathbf{a}}_i\Vert_2^2 ~ \text{s.t. } \Vert \mathbf{D} \mathbf{a} - \mathbf{b}_i \Vert_2 \leq \sqrt{m} \sigma_i \label{eq:noiseProjectionPrb}
\end{align}
where $\hat{\mathbf{a}}_{i}$ is the estimated HR SM row, $m = W H$. This projection enforces HR SM to be consistent with measured LR SM within an error bound determined by noise levels. Here, we implement the DC projection efficiently by a closed-form analytical solution (see Appendix): 
\begin{align}
\Tilde{\mathbf{b}}_i &= \mathbf{b}_i - \mathbf{D} \Tilde{\mathbf{a}}_i \\
\hat{\mathbf{a}}_{i} &=
\begin{cases}
    \Tilde{\mathbf{a}}_i, & \text{if } \Vert\Tilde{\mathbf{b}}_i \Vert_2\leq \sqrt{m} \sigma_i\\
    \Tilde{\mathbf{a}}_i + \left(1 - \frac{\sqrt{m} \sigma_i}{\Vert\Tilde{\mathbf{b}}_i \Vert_2}\right) \mathbf{D}^T \Tilde{\mathbf{b}}_i,              & \text{otherwise.}
\end{cases} 
\end{align}

\textbf{Network Loss:} The overall network can be cast as a feed-forward function $f_{\theta}(\mathbf{b}_i) = \hat{\mathbf{a}}_{i}$, where $\theta$ denotes the set of learnable parameters. The network is trained using $\ell_1$-norm loss between the recovered and reference HR SM rows:
\begin{align}
    \arg\min_{\theta} \sum_i \left\Vert f_{\theta}(\mathbf{b}_i) - \mathbf{a}_i \right\Vert_1
\end{align}
%

\begin{figure}[t]
	\centering
	\includegraphics[height=34mm]{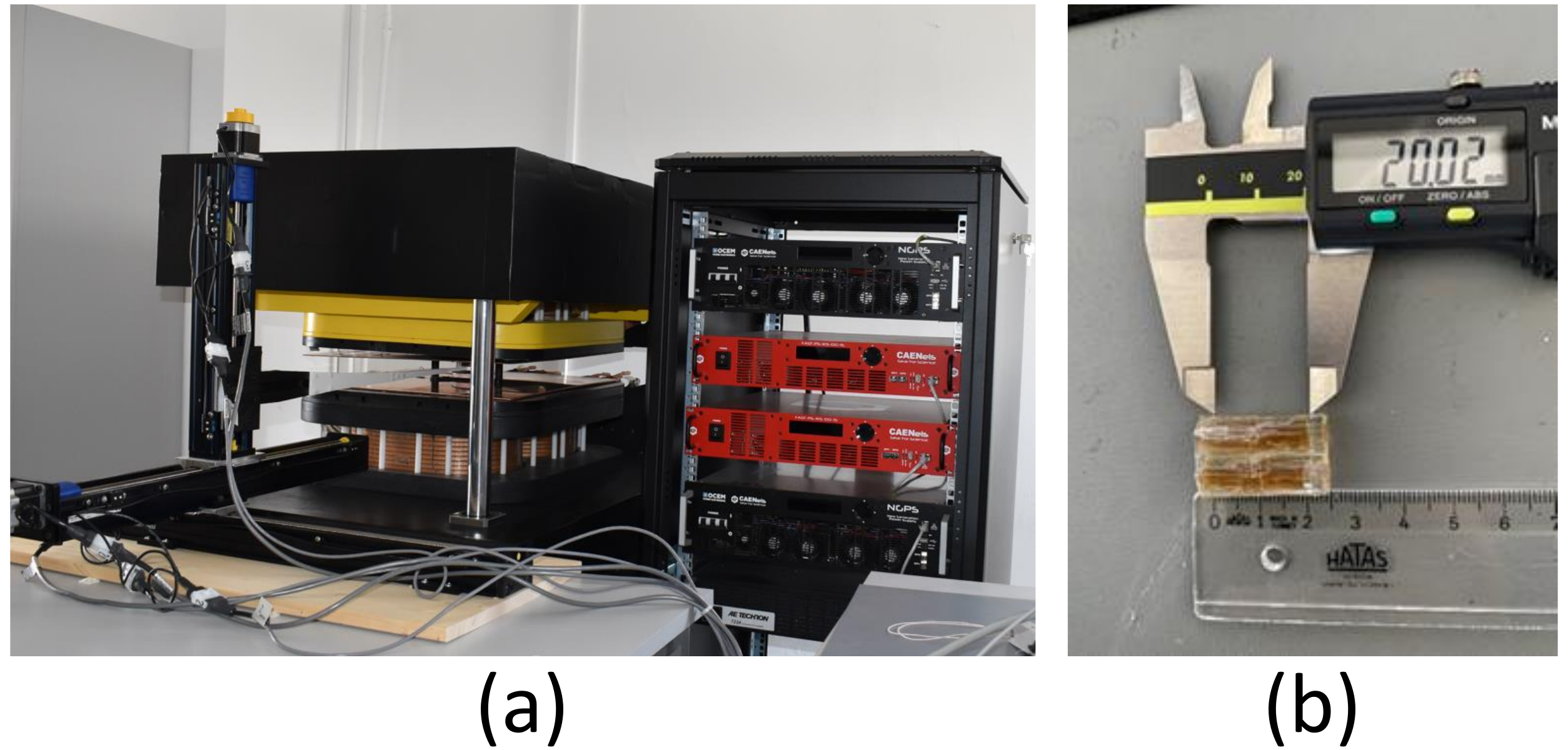}
	\caption{(a) In-house open-sided FFL scanner. (b) The imaging phantom used to acquire in-house MPI data. Cylindrical tubes were prepared with Perimag (Micromod GmbH, Germany) MNPs at a dilution ratio of 1:20 (405 µg/ml Fe). The length, inner radius and outer radius of the tubes were 20 mm, 2 mm and 4 mm, respectively. 
	}
	\label{fig:inhouse_phantom}
\end{figure}

\vspace{-0.2cm}
\section{Methods}

\subsection{Datasets}
\label{sec:datasets}
As experimental datasets are inherently noisy and of practically limited size, we first conducted controlled analyses on a simulated dataset. These analyses included ablation studies to demonstrate the components of the network architecture, ablation studies to assess the influence of the size and diversity of the training set, as well as comparisons to competing methods in noisy and noise-free settings. Next, we conducted analyses on two experimental datasets to demonstrate TranSMS on both FFP and FFL scanners. These analyses aimed to examine reliability against variability in SF strength, FOV position and MNP tracer type, as well as variability over time. In this section, we describe the details regarding the calibration measurements (Table~\ref{tab:datasetParameters}). The voxel size of each LR SM is $S \times S$ times larger compared to the respective HR SM.

\begin{table}[t]
\centering
\caption{Imaging protocols for HR SM in analyzed datasets.}
\label{tab:datasetParameters}
\resizebox{1\columnwidth}{!}{%
\begin{tabular}{|l|c|c|c|}
\hline
& \textbf{\begin{tabular}[c]{@{}c@{}}Simulated \\ Dataset \end{tabular}} & \textbf{\begin{tabular}[c]{@{}c@{}}Open MPI \\ Dataset (\#7) / \\  (\#8, \#9, \#10) \end{tabular}} & \textbf{\begin{tabular}[c]{@{}c@{}}In-house\\ MPI Dataset\end{tabular}} \\
\hline
\textbf{FOV Size (mm)}       & 32$\times$32$\times$1 & \begin{tabular}{c} 37$\times$37$\times$18.5 / \\ 66$\times$66$\times$27 \end{tabular}   & 32$\times$64$\times$2 \\
\hline
\textbf{Grid Size (px)}           & 32$\times$32$\times$1 &   \begin{tabular}{c} 37$\times$37$\times$37 / \\ 33$\times$33$\times$27 \end{tabular}  & 16$\times$32$\times$1 \\
\hline
\textbf{Sequence}       & FFL - 3$^\circ$ & FFP - Lissajous &  FFL - 3$^\circ$ \\
\hline
\textbf{MNP}            & [14.1-34.47] nm &
\begin{tabular}{c}Synomag / \\ Perimag \end{tabular} & Perimag
 \\
\hline
\textbf{\begin{tabular}[l]{@{}l@{}}Sample Sz.  (mm)  \end{tabular}}          & 1$\times$1$\times$1 & 2$\times$2$\times$1 & 2$\times$2$\times$2
 \\
\hline
\textbf{\begin{tabular}{c}SF Grad. \\ (T/m)\end{tabular}}    & \begin{tabular}{c}[0.4-1.03] \\ $\perp$ to FFL \end{tabular}   &  \begin{tabular}{c}-1$\times$-1$\times$2 / \\ -0.5$\times$-0.5$\times$1\end{tabular} &  \begin{tabular}{c}[0.3-0.6]\\ $\perp$ to FFL \end{tabular} \\
\hline
\textbf{DF Freq.} & 23.2 kHz  &  \begin{tabular}[c]{@{}c@{}}2.5/102 MHz\\ 2.5/96 MHz\\ 2.5/99 MHz\end{tabular}  &  23.2 kHz
 \\
\hline
\textbf{DF Amp. (mT)}  &   0$\times$0$\times$[6.4-16] & 12$\times$12$\times$12 & 0$\times$0$\times$9 \\
\hline
\end{tabular}
}
\end{table}

\subsubsection{Simulated Dataset}
We used an in-house MPI simulator modeling the open-sided FFL scanner topology in \cite{cbtop2020}. A 32$\times$32 mm$^2$ FOV was scanned with a sample of size 1$\times$1 mm$^2$ by rotating the FFL in the x-y plane. MNP saturation magnetization and temperature were 0.55 $T/\mu_0 $ and 300 $^\circ$K. A sinusoidal DF with varying amplitude (depending on the SF gradient and FOV size) was applied in the z-direction. \textcolor{black}{The received signal was sampled at 5 MS/s for a receive coil with homogeneous sensitivity in the z-direction}. Following Fourier transformation of time-domain signals, the components corresponding to harmonics 2-to-9 were captured. Finally, imaging data were generated for a numerical phantom.

Simulated SMs were split to use 66 for training, 30 for validation and 34 for testing. \textcolor{black}{The validation set contained SF gradients [0.70:0.265:1.03] T/m, and MNP diameters [15.17:2.145:34.47] nm. For the training and test sets, 100 SMs were generated with SF gradients of [0.40:0.075:1.00] T/m, MNP diameters of [14.10:2.145:33.40] nm. For the test set, we exclusively reserved 19 of these SMs with SF gradient 1 T/m or MNP diameter 33.40 nm, and \textcolor{black}{15} additional randomly selected SMs were added for improving diversity.} 

\subsubsection{Open MPI Dataset}
Open MPI is a public dataset with calibration and imaging measurements on a preclinical FFP scanner (Bruker, Ettlingen) \cite{openmpi}. We analyzed calibration data collected at 37$\times$37$\times$18.5 mm$^3$ and 66$\times$66$\times$27 mm$^3$ FOVs, a sample of size 2$\times$2$\times$1 mm$^3$ and a 3D Lissajous sequence. Signals from 3 coils were sampled at 2.5 MS/s. SM rows with high-SNR responses were selected (SNR$>$5) \cite{otherReconMPI, deepImagePriorMPI, superResolutionMPISM}, and whitened based on the noise power computed using the provided SNR levels. Noise was estimated to induce an intrinsic 3.38\% normalized root-mean-squared-error (nRMSE), serving as a performance bound for SM recovery. Imaging data collected from a Perimag-filled resolution phantom for [-0.5, -0.5, 1] T/m gradients were used. \textcolor{black}{To assess generalizability across time, longitudinally measured SMs with identical imaging protocols albeit spatially-misaligned FOVs were used to construct the training (\#8, \#9) and test (\#10) sets. To assess generalizability across MNP type and SF gradients, an SM sampled using Synomag-D (Micromod GmbH, Germany) and [-1, -1, 2] T/m SF gradients was reserved for training (\#7), and another SM using Perimag (Micromod GmbH, Germany) and [-0.5, -0.5, 1] T/m SF gradients was reserved for testing (\#10). SM volumes were split into separate slices.}

\begin{table}
\centering
\caption{Average nRMSE ($\%$) in HR SM recovery for ablation studies performed under noisy setting at $2\times$-$8\times$ SR.} 
\label{table:comparison_smNsyAblation}
\begin{tabular}{|c|c|c|c|} 
\hline 
& $2\times$ & $4\times$ & $8\times$ \\ 
\hline 
RDSR & 0.84\% & 1.18\% & 4.32\% \\
\hline 
CTSR & 0.45\% & 0.70\% & 4.06\% \\ 
\hline
CNN-ViT & 0.67\% & 0.81\% & 3.91\% \\
\hline 
RD-ViT & 0.57\% & 0.76\% & 4.14\% \\
\hline 
CNN-CT & 0.51\% & 0.68\% & 4.31\% \\
\hline
TranSMS & \textbf{0.41\%} & \textbf{0.58\%} & \textbf{3.71\%} \\ 
\hline
\end{tabular}
\end{table}

\subsubsection{In-house MPI Dataset}

Data were collected on the open-sided ASELSAN FFL scanner (Fig. \ref{fig:inhouse_phantom}a)\cite{cbtop2020}. A 32$\times$64 mm$^2$ FOV was scanned with an undiluted Perimag sample, with a sample of size 2$\times$2 mm$^2$ by rotating the FFL in the x-y plane. SF gradients of $[0.3, 0.4, 0.5, 0.6]$ T/m were applied, with 9 mT DF amplitude for 10 ms per angle. A 10\% duty cycle was used to prevent drive coil overheating. The received signal was amplified (gain = 5) and filtered (cutoffs: 10-300 kHz) on an SR-560 pre-amplifier (SRS, MA, USA), and sampled at 5 MS/s. Frequency components around harmonics $2$-to-$11$ were selected with a $500$ Hz bandwidth. High-SNR rows were selected (SNR$>$5) and whitened. Noise was estimated to induce an intrinsic 4.36\% nRMSE. For retrospective analyses that assessed generalizability across SF gradient strength, \textcolor{black}{the training set contained 3 SMs at [$0.3, 0.4, 0.6$] T/m, the test set contained 1 SM at $0.5$ T/m}. To improve reconstruction quality, SMs were cropped to the sensitive region of the receive coil ($13 \times 23$ grid size). \textcolor{black}{For a prospective analysis to demonstrate the full SNR benefits of TranSMS, we separately measured an LR SM using a larger 4$\times$4 mm$^2$ MNP sample, and super-resolved HR SM using the retrospectively trained model.}
Imaging data were collected by applying a 0.5 T/m SF gradient to cylindrical tube (inner radius: 2 mm, outer radius: 4 mm) phantoms filled with diluted Perimag (Fig. \ref{fig:inhouse_phantom}b). \textcolor{black}{In retrospective experiments, a phantom with two 20-mm-long tubes with 8-12 mm inter-tube spacing was used.}
\textcolor{black}{In prospective experiments, a 23-mm-long tube, and a 21-mm-long tube with a 5-mm-long central stenosis region of 1-mm inner radius with 6-7 mm inter-tube spacing was used.}

\subsection{TranSMS Implementation}
Hyperparameters including number of network layers, number of channels, and learning rate were selected via one-fold cross-validation on the simulated dataset. Non-overlapping training, validation and test sets were constructed. Parameters were optimized separately for each SR factor ($2\times$, $4\times$, $8\times$) to maximize validation performance in terms of nRMSE. The common set of parameters among SR factors included $n_{RDB} = 4, C_C = 24, n_{GR} = 6, C_T = 64, n_A = 4, C_{cat} = 48$. Meanwhile, $C_1 = [24, 24, 64]$ and $n_{CL} = [5, 8, 9]$ were selected for $2\times$, $4\times$, and $8\times$ SR, respectively. The transformer module contained $n_{TB} = 3$ stages, where the first two used convolutions of stride $1$, and the last stage used stride $2$. MLP blocks used Gaussian error linear unit (GELU) activation functions, RDBs used rectified linear unit (ReLU) activation functions, and upsampler blocks used leaky ReLU.

\subsection{Competing Methods}
TranSMS was comparatively demonstrated against several state-of-the-art techniques for SM calibration including traditional, CS-based and SR-based methods. 

\subsubsection{Bicubic \& Strided Bicubic} As a traditional method, bicubic interpolation was used to upsample LR SM \cite{Gungor_iwmpi}. Two variants were implemented based on LR SM measurements with large MNP sample size versus strided subsampling of HR SM measurements with small MNP sample size. Real and imaginary parts of SM were treated separately.

\subsubsection{Compressed Sensing} HR SM measurements were randomly undersampled in 2D, and data were assumed to be sparse in the Fourier domain \cite{Weber_2015}. An ADMM-based algorithm was used to solve the CS optimization problem \cite{FastCalibrationSilbey}. To prevent edge artifacts, the FOV was extended by $4$ pixels in all directions, the original FOV was retained in recovered SM. 

\subsubsection{SRCNN} For SR-based recovery, LR SM rows were upsampled with the popular SRCNN model \cite{reviewPaperSR, srcnn}. SRCNN performs bicubic upsampling followed by projection through 3 cascaded convolutional layers. Real and imaginary parts of SM were provided as separate channels at input and output. 

\subsubsection{VDSR} As another SR method, LR SM rows were upsampled with VDSR \cite{vdsr}. VDSR performs bicubic upsampling followed by 20 cascaded convolution layers to better preserve high-frequency information. Real and imaginary parts of SM were provided as separate channels at input and output. 

\subsubsection{2d-SMRnet} Finally, the 3d-SMRnet model in \cite{2d-SMRnet} was adapted for 2D SR-based calibration by converting convolution kernels from 3D to 2D. During calibration, 2d-SMRnet performs strided LR SM measurements with a small MNP sample. Real and imaginary parts of SM were provided as separate channels at the model input and output.

\begin{figure}[t]
\centering
	\includegraphics[width=0.9\columnwidth]{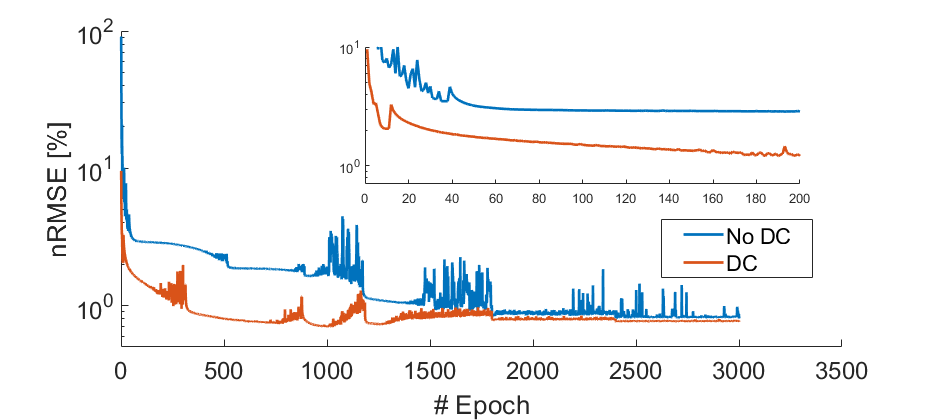}
\caption{Validation performance of TranSMS with (red curve) and without (blue curve) the data-consistency (DC) module. The HR SM recovery performance is reported as nRMSE across training epochs. The upper-right panel displays a zoomed-in portion of nRMSE curves during the initial stages of training. }
	\label{fig:DCmodule}
\end{figure}

\begin{figure}[t]
\centering
	\includegraphics[width=0.9\columnwidth]{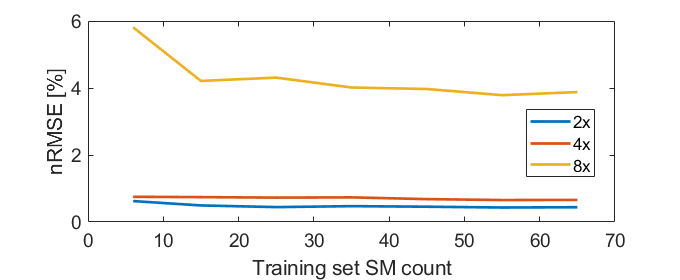}
\caption{SM recovery performance of TranSMS as a function of the number of SMs included in the training set for SR factors of $2\times$-$8\times$.}
	\label{fig:trainingSetAbl}
\end{figure}

\subsection{Analysis Procedures}

\subsubsection{SM Recovery}
DL models were implemented using the PyTorch framework, and trained via the Adam optimizer. For TranSMS, the learning rate was halved at every $N_{epoch}$/5 epochs, where $N_{epoch}$ is number of epochs. For 2d-SMRnet, SRCNN and VDSR, learning rate schedules proposed in the respective manuscripts were adopted \cite{2d-SMRnet, srcnn, vdsr}. The weight decay parameter was set to $10^{-8}$ for simulated, and to $0$ for experimental datasets as in \cite{2d-SMRnet}. The learning rate was selected for each dataset and SR factor via cross-validated search across the range $[10^{-8}, 1]$. Bicubic and CS methods were implemented in MATLAB (MathWorks, Natick USA). Bicubic interpolation assumed default settings in the \textit{imresize} function. In CS, intensity of each whitened SM row was normalized separately for numerical stability, and restored to its original scale following recovery. Step size $\mu = 1$ and \# of iterations = $1000$ were selected via cross-validation. The bound on the noise level for the constrained problem was estimated from background measurements.

Methods were compared in super-resolving LR SM. Background measurements were used to estimate $\sigma_i$ in Eq.~\eqref{eq:noiseProjectionPrb}, and to perform background subtraction. \textcolor{black}{In the simulated dataset, LR SM was simulated for a large MNP sample spanning an LR voxel.} Gaussian noise was added to maintain 30 dB SNR \cite{cbtop2020}. Networks were trained to estimate noise-free HR SM given noisy LR SM. We assessed performance at $2\times$--$8\times$ SR, reducing calibration time by $4$--$64$-fold, respectively. 

In experimental datasets, \textcolor{black}{independent LR and HR SM measurements can carry inconsistencies due to nuisance variability in the system (e.g., sample misalignment, temperature drift), biasing trained models and degrading recovery performance. To alleviate this issue, LR SMs were generated retrospectively using the linearity of the MPI signal, through box-car downsampling of HR SMs to permit unbiased reporting of performance metrics}. Training sets were expanded via data augmentation by horizontal and vertical flipping of SM rows. Networks were trained to estimate measured HR SM given LR SM, considering $2\times$-$8\times$ SR.  \textcolor{black}{To prospectively demonstrate TranSMS, an LR SM measured on the in-house scanner using a large MNP sample spanning an LR voxel was super-resolved using the retrospectively trained model.}

\begin{table}
\centering
\caption{Average nRMSE ($\%$) in HR SM recovery for the simulated dataset under (noiseless/noisy) settings at $2\times$-$8\times$ SR. } 
\label{table:comparison_smNsless}
\resizebox{1\columnwidth}{!}{%
\begin{tabular}{|c|c|c|c|} 
\hline 
 & $2\times$ & $4\times$ & $8\times$  \\ 
\hline 
Bicubic & 3.53\% / 4.02\% & 14.82\% / 14.95\% & 46.88\% / 46.92\% \\
\hline 
Bicub. (str.) & 14.9\% / 16.6\% & 42.61\% / 50.20\% & 80.20\% / 93.34\% \\
\hline 
CS & 4.15\% / 9.81\% & 23.58\% / 37.94\% & 80.60\% / 101.39\% \\
\hline 
SRCNN & 33.5\% / 38.8\% & 36.48\% / 35.84\% & 53.93\% / 50.00\% \\
\hline 
VDSR & 2.66\% / 6.51\% & 5.49\% / 13.36\% & 20.02\% / 48.16\% \\
\hline 
2d-SMRnet & 0.12\% / 1.34\% & 0.36\% / 5.30\% & 1.11\% / 64.02\% \\
\hline 
TranSMS & \textbf{0.09\% / 0.41\%} & \textbf{0.13\% / 0.58\%} & \textbf{0.40\% / 3.71\%} \\ 
\hline
\end{tabular} 
}
\end{table}

\subsubsection{Image Reconstruction}
Recovered HR SMs were used to perform image reconstruction. Among advanced methods \cite{deepImagePriorMPI, otherReconMPI, Knopp_2021_image_recon}, we opted for an ADMM-based algorithm with fast convergence implemented in PyTorch \cite{ilbey2017comparison,boydADMM}. Data were whitened using background measurements prior to reconstruction, so $\epsilon$ was set to $\sqrt{M}$ where $M$ is the number of HR SM rows. Reconstruction using the reference HR SM was taken as as a benchmark for performance evaluations. Hyperparameters were selected via cross-validated search to optimize quality of the reference image in the simulated dataset. The selected parameters were $\alpha_1=0.95$, $\alpha_{TV} = 0.05$, step size of $\mu=10$, and $5000$ iterations.

\subsection{Quantitative Assessments}
Performances in HR SM recovery and image reconstruction were assessed via quantitative metrics. Assuming $\hat{\mathbf{A}}$ is the recovered HR SM and $\mathbf{A}$ is the reference HR SM, the data-consistency constraint in Eq.~\eqref{eq:imageRecon} can be expressed as:
\begin{align}
    \Vert \hat{\mathbf{A}} \mathbf{x} - \mathbf{y} \Vert_2^2 = \Vert \mathbf{R} \mathbf{x} \Vert_2^2 + \Vert \mathbf{A} \mathbf{x} - \mathbf{y} \Vert_2^2 + 2 \mathbf{x}^T \mathbf{R}^T \left( \mathbf{A} \mathbf{x} - \mathbf{y}\right)
\end{align}
where $\mathbf{R} = \hat{\mathbf{A}} - \mathbf{A}$ denotes the residual error in HR SM recovery. Given zero-mean measurement noise, the expected value of $\mathbf{A} \mathbf{x} - \mathbf{y}$ is 0. Data-fidelity error is then proportional to the Frobenius norm of the residual error, $\Vert \mathbf{R}\Vert_F^2$. Thus, we measured the quality of HR SM recovery via nRMSE:
\begin{align}
    nRMSE(\hat{\mathbf{A}}) &= \Vert \hat{\mathbf{A}} - \mathbf{A} \Vert_F / \Vert \mathbf{A} \Vert_F  \label{eq:nRMSE}
\end{align}
For image reconstruction, we adopted the popular peak Signal-to-Noise-Ratio (pSNR). Assuming $\mathbf{x}_{ref}$ is the reference image:
\begin{align}
    pSNR(\mathbf{x}) = 20 \log_{10} &\left(\sqrt{N} \Vert \mathbf{x}_{ref}\Vert_{\infty} / \Vert  \mathbf{x} - \mathbf{x}_{ref}\Vert_2 \right)
\end{align}

\section{Results}

\begin{table}
\centering
\caption{Average pSNR (dB) in image reconstruction for the simulated dataset under noisy setting at $2\times$-$8\times$ SR.} 
\label{table:comparison_imag30}
\begin{tabular}{|c|c|c|c|} 
\hline 
& $2\times$ & $4\times$ & $8\times$  \\ 
\hline
Bicubic  & 24.85  & 14.05  & 9.70 \\
\hline
Bicubic (str.)  & 15.00  & 9.86  & 8.36 \\
\hline
CS  & 19.92  & -4.15  & -14.20 \\
\hline
SRCNN  & 16.56  & 12.39  & -8.25 \\
\hline
VDSR  & 23.07  & 17.32  & -9.62 \\
\hline
2d-SMRnet  & 24.77  & 20.38  & -11.60 \\
\hline
TranSMS  & \textbf{26.35}  & \textbf{25.72}  & \textbf{20.89} \\
\hline
\end{tabular}
\end{table}

\subsection{Ablation Studies}

\subsubsection{Network Ablation}

We first demonstrated the role of each \textcolor{black}{network component in} TranSMS. Analyses were performed on the simulated dataset where a noise-free ground truth is available for absolute performance evaluation. First, we compared TranSMS against variants where \textcolor{black}{the convolutional transformer (CT) module was ablated (residual-dense super-resolution; RDSR) or the residual-dense (RD) convolutional module was ablated (convolutional transformer super-resolution; CTSR).} Second, we evaluated the specific implementations of transformer and convolutional modules. \textcolor{black}{To evaluate the benefit of CT blocks, we implemented a variant containing vanilla vision transformer (ViT) modules (residual-dense vision transformer; RD-ViT). To evaluate the benefit of RD blocks, we implemented a variant with vanilla CNN modules (CNN with convolutional transformer, CNN-CT). The combination of vanilla CNN and vanilla transformer was also considered (CNN with vision transformer, CNN-ViT).} Table~\ref{table:comparison_smNsyAblation} lists performance for TranSMS and its variants. \textcolor{black}{Overall, TranSMS yields higher performance than both RDSR and CTSR, indicating the importance of its transformer and convolutional modules. TranSMS outperforms outperforms RD-ViT indicating the benefit of CT over ViT, and it also outperforms CNN-CT indicating the benefit of RD over CNN.} Finally, we evaluated the benefit of the DC module by comparing TranSMS variants with and without DC projection. Fig.~\ref{fig:DCmodule} plots nRMSE for $4\times$ SR, showing that the DC module enables faster and more stable convergence.

\begin{figure}[t]
    \centerline{\includegraphics[width=1.01\columnwidth]{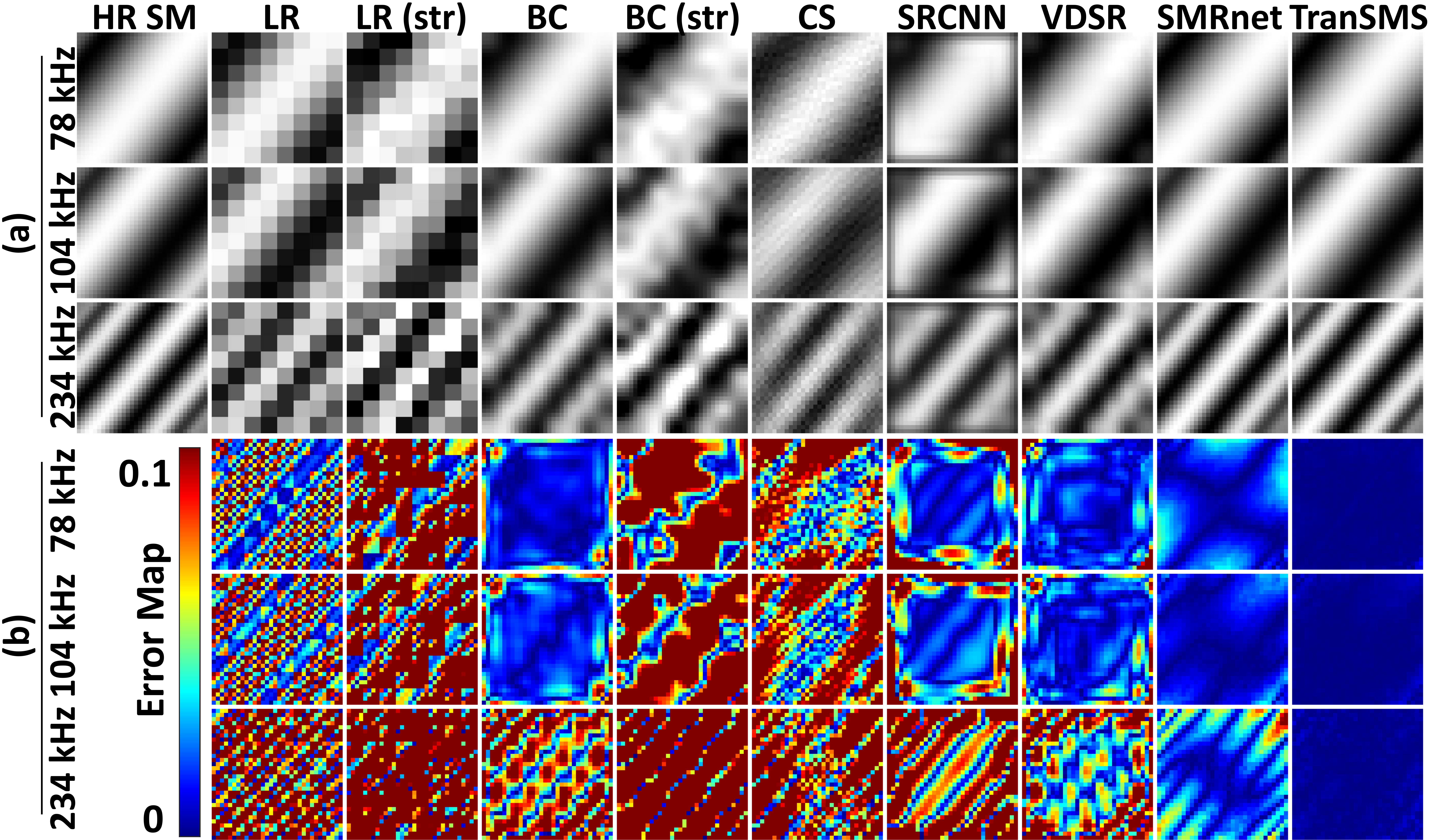}}
\caption{HR SM recovery in the simulated dataset for $4\times$ SR using noisy measurements. SF gradient strength and MNP size differed between the training and test sets. SM rows corresponding to different harmonics at 45$^\circ$ are shown. \textbf{(a)} Recovered HR SMs are displayed along with the input LR SM, and input strided (str) LR SM. Strided bicubic (BC) interpolation and 2d-SMRnet receive as input the strided LR SM. \textbf{(b)} Respective error maps (see colorbar).
}
	\label{fig:sm1Out4x}
\end{figure}

\begin{figure}[t]
 	\centerline{\includegraphics[width=\columnwidth]{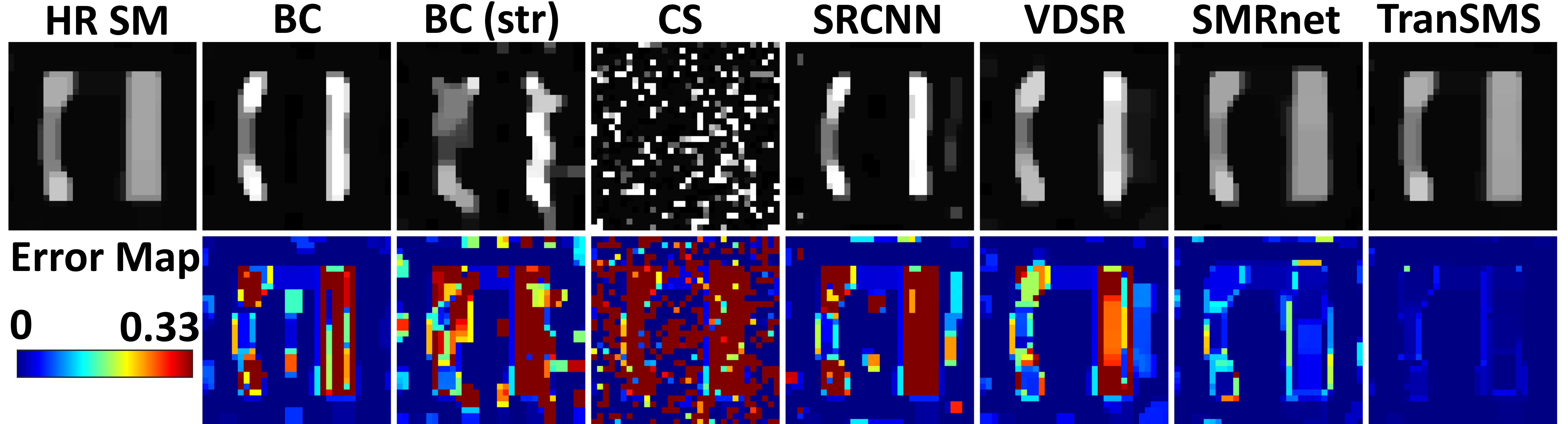}}
\caption{Phantom images from the simulated dataset at 1 T/m SF gradient strength reconstructed with recovered HR SMs at $4\times$SR are displayed along with the reference reconstruction using the reference HR SM (top row). Error maps for each method are also displayed (bottom row). The phantom comprised two 22-mm long tubes, the right tube's width fixed at 6 mm, and the left tube's width ranging in 2-6 mm. Inter-tube spacing varied in 8-12 mm. }
	\label{fig:imagesNo3Out_scale_4}
\end{figure}

\subsubsection{Training Set Ablation}

Next, we assessed the influence of the number and diversity of the SMs in the training set on model performance. \textcolor{black}{These analyses were conducted on the simulated dataset.} First, we varied the number of SMs used for training in the range [6 66] via random selection, while the original test set was maintained. Figure \ref{fig:trainingSetAbl} shows nRMSE in SM recovery as a function of the number of training SMs. Near-optimal results are achieved around 15 SMs with diminishing returns for larger training sets. Second, we compared two training sets with 6 SMs each, where one was restricted to a narrow range of SF and MNP parameters (gradient strength in T/m, diameter in nm): (0.3, 18.39), (0.3, 20.53), (0.3, 24.82), (0.7, 16.24), (0.7, 22.67), (0.7, 24.82). The other training set was randomly selected. \textcolor{black}{We find that the restricted and randomized training sets yield moderate performance differences, with an average nRMSE of 2.54\% for randomized and 3.42\% for restricted sets.}

\begin{figure}[t]
\centering
\includegraphics[scale=0.12]{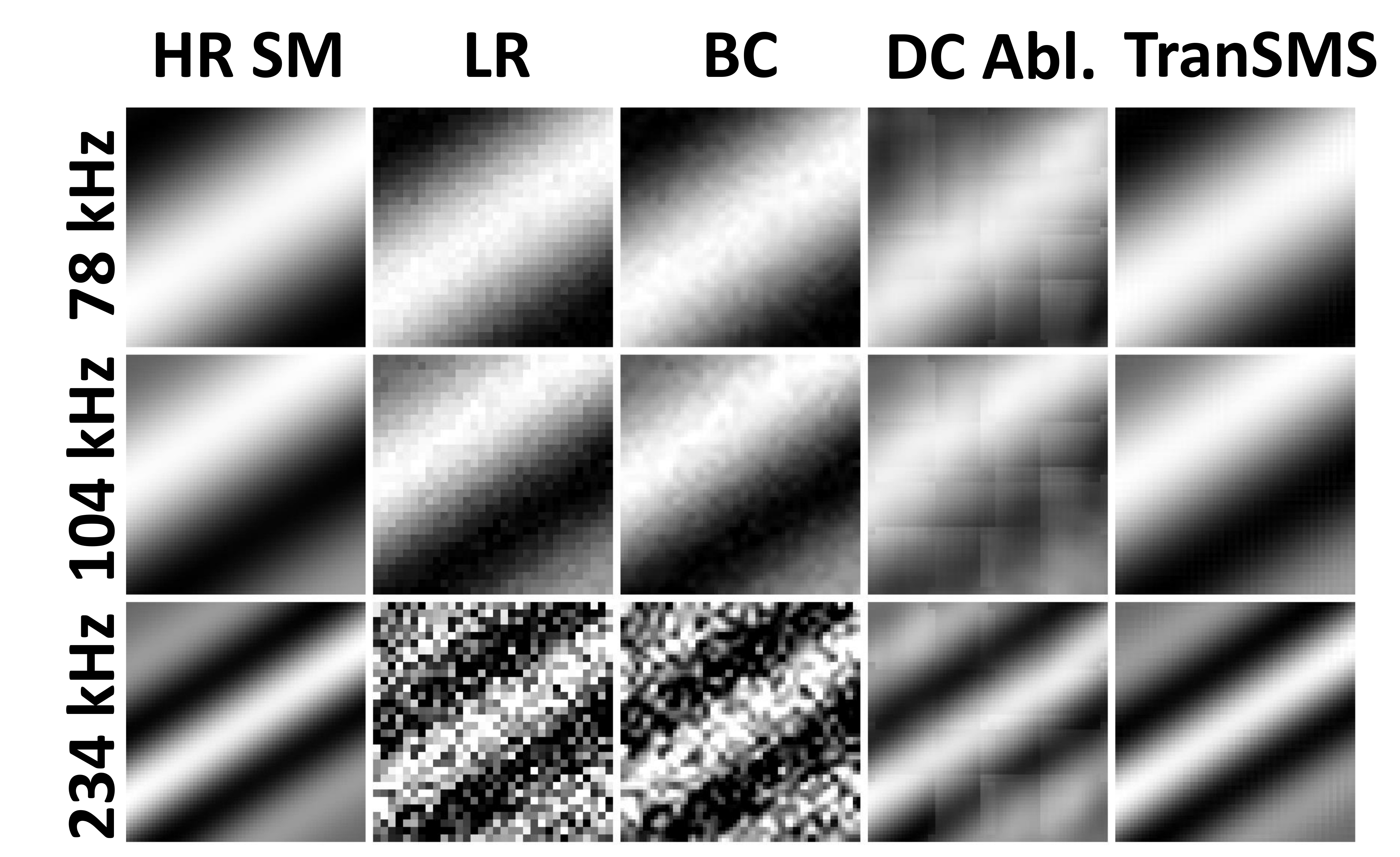}
	\caption{A pre-trained TranSMS model was used to super-resolve a 64$\times$64 HR SM by processing overlapping 16$\times$16 patches in the input 32$\times$32 LR SM. SM rows for different harmonics at 45$^\circ$ are shown. Recovered HR SMs are shown under 30 dB measurement SNR, along with the input LR SM and the reference HR SM.}
	\label{fig:myHRfig}
\end{figure}

\subsection{Analyses on Simulated Dataset}

\subsubsection{SF \& MNP Tracer Variability}

We first demonstrated TranSMS on the simulated dataset for HR SM recovery at $2\times$-$8\times$ SR. \textcolor{black}{We assessed reliability against variability in SF gradients and MNP diameter between training and test sets, under both noise-free and noisy settings.} Table \ref{table:comparison_smNsless} lists nRMSE of recovered HR SMs. TranSMS yields superior performance against competing methods in all cases, with 21.96\% lower nRMSE over the second-best method on average in noisy settings.  
Representative SM rows are displayed in Fig.~\ref{fig:sm1Out4x} for 4$\times$ SR. 
Bicubic interpolation and VDSR have visible blurring, CS shows prominent artifacts and noise, and SRCNN shows edge artifacts and blurring. While 2d-SMRnet is often the top contender, it has a high error rate at 8$\times$ in the noisy setting. This could be attributed to the relatively high complexity of 2d-SMRnet rendering it more susceptible to overfitting with noisy training data.

We then investigated the effects of recovered HR SMs on image reconstruction performance. Table~\ref{table:comparison_imag30} lists pSNR for data simulated at 30 dB SNR. TranSMS outperforms all competing methods with $6.04$ dB higher pSNR over the second-best method on average. Representative phantom images at 4$\times$ SR are shown in Fig.~\ref{fig:imagesNo3Out_scale_4}. DL methods outperform CS in terms of residual artifacts, and bicubic interpolation in terms of visual acuity. Among DL models, TranSMS offers the reconstruction with the highest similarity to the reference image, whereas 2d-SMRnet is the second best with occasional intensity errors. Methods successful in SM recovery generally yield higher downstream reconstruction performance.

\subsubsection{Super-Resolving Higher Grid-Size}
To assess the performance of TranSMS in recovering a higher grid size than the reference HR SM it has been trained to recover, we trained TranSMS to estimate 32$\times$32 HR SM given 16$\times$16 LR SM. We then used the trained model to super-resolve a 64$\times$64 HR SM by processing overlapping 16$\times$16 patches in a given 32$\times$32 SM. 
Representative results for recovered SMs at 30 dB SNR are shown in Fig. \ref{fig:myHRfig}. 
TranSMS yields highly similar SM estimates to the reference HR SM although it has not been trained using data at the target SM resolution. \textcolor{black}{Note that prominent block artifacts are evident when the DC module is ablated, suggesting the importance of explicit DC projections for out-of-domain generalization.}

\subsection{Analyses on Open MPI Dataset}

\begin{table}[t]
\caption{Average nRMSE ($\%$) in HR SM recovery for Open MPI at $2\times$-$8\times$ SR. Calibration scans were performed at different times and with unaligned FOVs for SMs in training and test sets. } 
\centering
\label{table:comparison_OpenMPI}
\centering
\begin{tabular}{|c|c|c|c|} 
\hline 
 & $2\times$ & $4\times$ & $8\times$  \\ 
\hline 
Bicubic & 4.55\% & 18.13\% & 52.02\% \\
\hline 
Bicubic (str.) & 16.86\% & 47.41\% & 92.08\% \\
\hline 
CS & 8.81\% & 51.48\% & 101.31\% \\
\hline 
SRCNN & 53.65\% & 53.16\% & 79.29\% \\
\hline 
VDSR & 4.55\% & 12.62\% & 39.19\% \\
\hline 
2d-SMRnet & 19.69\% & 48.94\% & 75.37\% \\
\hline 
TranSMS & \textbf{3.15\%} & \textbf{6.19\%} & \textbf{20.58\%} \\
\hline 
\end{tabular}
\end{table}

\begin{table}[t]
\centering
\caption{Average nRMSE ($\%$) in HR SM recovery / pSNR (dB) in image reconstruction for Open MPI at $2\times$-$8\times$ SR. SF gradient strength and MNP tracer differed for SMs in training and test sets.} 
\label{table:comparison_OpenMPISMAndimageRecon}
\begin{tabular}{|c|c|c|c|} 
\hline 
& $2\times$ & $4\times$ & $8\times$ \\ 
\hline 
Bicubic & 4.55\% / 34.77  & 18.13\% / 17.15  & \textbf{52.02\%} / 11.34 \\
\hline
Bicubic (str.) & 16.86\% / 33.30 & 47.41\% / 18.62 &  92.08\% / 13.88 \\
\hline
CS & 8.81\% / 33.89 & 51.48\% / 20.47  & 101.31\% / \textbf{16.12} \\
\hline
SRCNN & 50.88\% / 32.21  & 62.81\% / 19.39 & 106.76\% / -- \\
\hline
VDSR & 3.34\% / 38.23  & 11.83\% / 30.59  &  113.81\% / 10.01 \\
\hline
2d-SMRnet & 6.85\% / 33.79  & 17.22\% / 24.81 & 78.88\% / 12.60 \\
\hline
TranSMS & \textbf{3.32\% / 38.54}  & \textbf{10.66\% / 31.96}  &  114.45\% / 13.38 \\
\hline
\end{tabular}
\end{table}

\subsubsection{Temporal \& FOV Variability}

\textcolor{black}{Next, we demonstrated TranSMS for HR SM recovery in Open MPI using two separate analyses with different training sets albeit a common test set}. \textcolor{black}{First, we assessed reliability against variability across time and FOV positioning between the training and test SMs. Table~\ref{table:comparison_OpenMPI} lists nRMSE when the training and test sets contained SMs measured with the same protocol albeit at different time points and with unaligned FOVs.} TranSMS yields superior performance against competing methods, with 8.81\% lower nRMSE on average over the closest competitor.

\subsubsection{SF \& MNP Tracer Variability}

\textcolor{black}{We also assessed reliability against variability in SF gradients and MNP tracer.} Table~\ref{table:comparison_OpenMPISMAndimageRecon} lists respective nRMSE in HR SM recovery when the training and test SMs differed in their SF and MNP parameters. TranSMS performs better than competing methods, except for $8\times$ SR where all methods perform poorly and bicubic interpolation yields the lowest nRMSE. TranSMS lowers nRMSE by 0.60\% over the second-best method across $2\times$-$4\times$ SR. 

\begin{figure}[t]
	\centerline{\includegraphics[width=1.01\columnwidth]{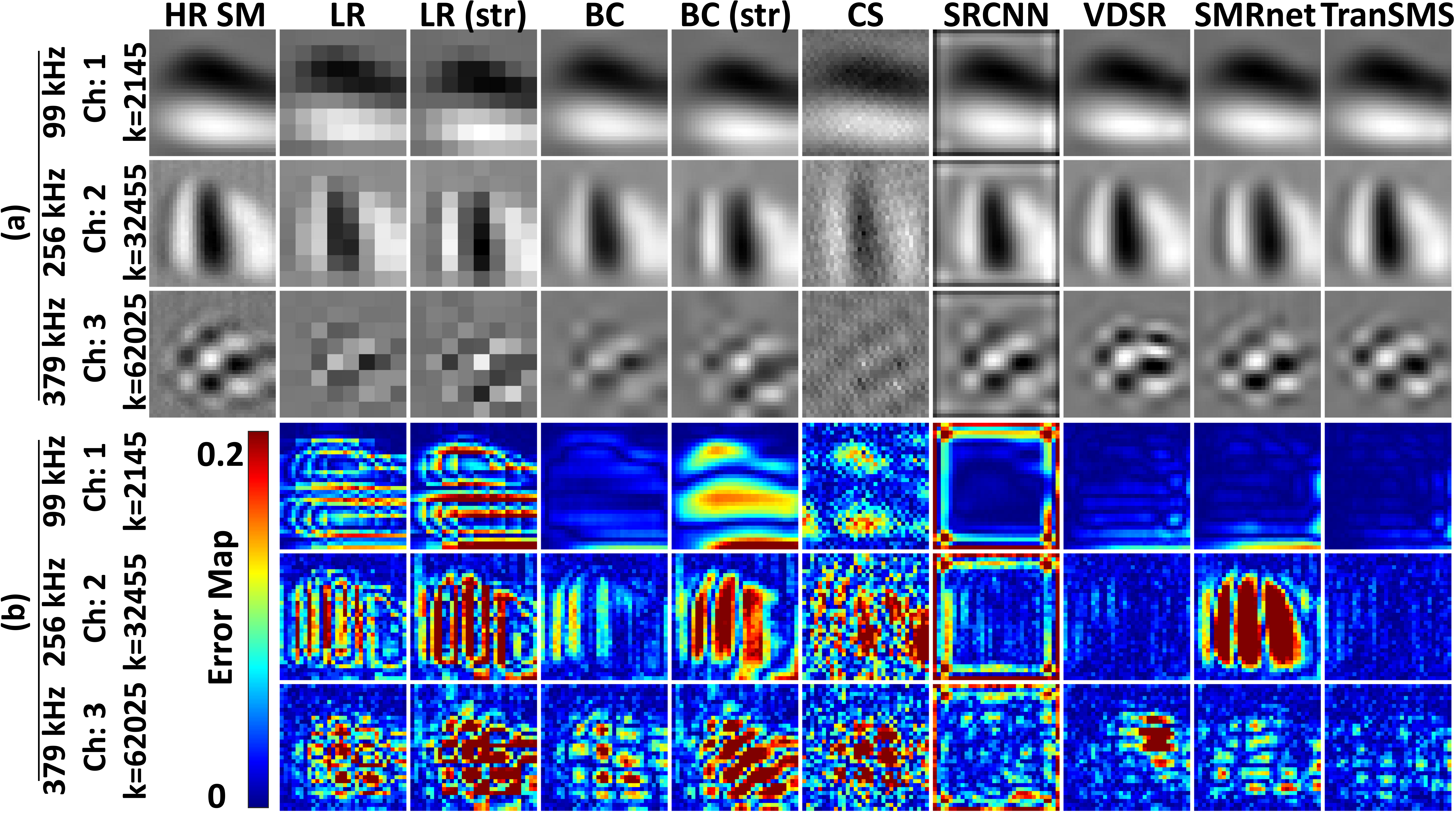}}
\caption{HR SM recovery in Open MPI for $4\times$ SR. Both the SF gradient strength and MNP tracer differed for SMs in training and test sets. The SM rows corresponding to the center slice at different frequency components and receive channels (Ch) are shown (k denotes the respective SM row index). \textbf{(a)} Recovered HR SMs and \textbf{(b)} respective error maps are displayed along with the input LR SM, input strided LR SM and reference HR SM.
}
	\label{fig:openMPI4x}
\end{figure}

\begin{figure}[t]
\centering
	\centerline{\includegraphics[width=\columnwidth]{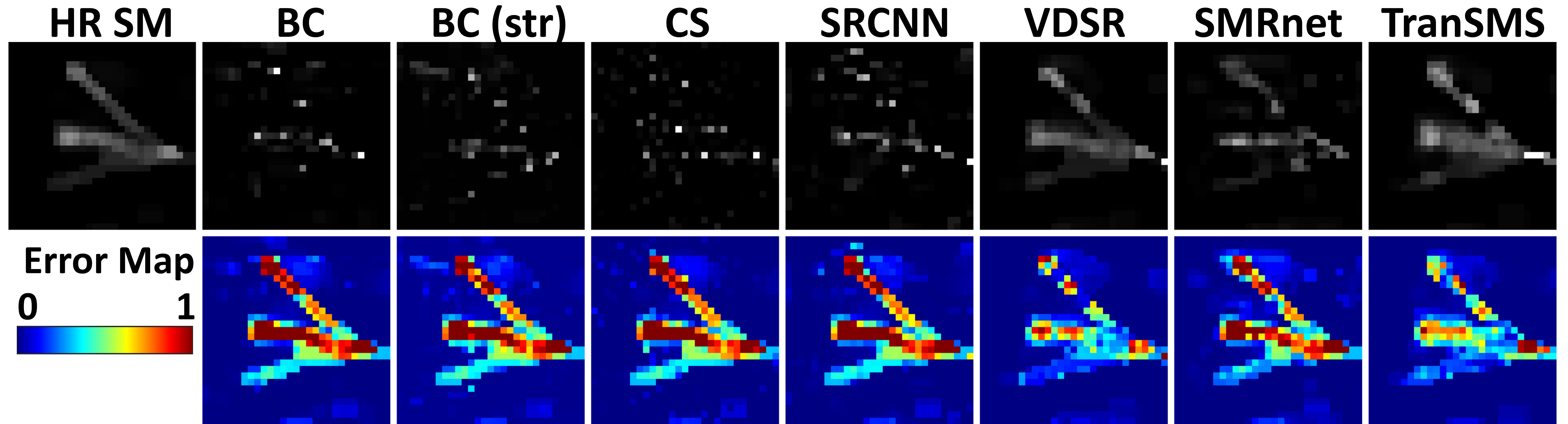}}
\caption{Image reconstruction in Open MPI for $4\times$ SR. Phantom images reconstructed with recovered HR SMs are displayed along with the benchmark reconstruction using the reference HR SM (top row). Respective error maps are also displayed (bottom row).
}
	\label{fig:openMPIimage}
\end{figure}

Recovered HR SMs at 4$\times$ SR are displayed in Fig.~\ref{fig:openMPI4x}. TranSMS generates a more similar estimate to the reference HR SM than competing methods. Bicubic interpolation suffers from visible blurring, and CS shows artifacts. SRCNN performs poorly with visible boundary artifacts. While 2d-SMRnet achieves high spatial acuity, it shows some geometric warping. VDSR yields a geometrically accurate estimate with some intensity errors. In contrast, TranSMS yields an HR SM of high spatial acuity and consistency to the reference HR SM. Table~\ref{table:comparison_OpenMPISMAndimageRecon} lists pSNR for subsequent image reconstructions except for SRCNN that did not converge and returned an empty image at $8\times$ SR. On average, TranSMS outperforms the closest competitor by 1.67 dB pSNR. \textcolor{black}{A representative cross-section from the reconstructed phantom at 4$\times$ SR is shown in Fig.~\ref{fig:openMPIimage}, where methods with successful SM recovery demonstrate improved reconstruction results.}

\begin{figure*}[t]
    \begin{minipage}{0.4\textwidth}
\caption{HR SM recovery in the in-house MPI dataset for $4\times$ SR. The SM rows corresponding to different harmonics at 81$^\circ$ are shown. \textbf{(a)} Recovered HR SMs and \textbf{(b)} respective error maps for each competing method are displayed along with the input LR SM, input strided LR SM and reference HR SM.
}
	\label{fig:aselsanFFL8}
	\end{minipage}
	\begin{minipage}{0.6\textwidth}
	\centerline{\includegraphics[width=0.99\textwidth]{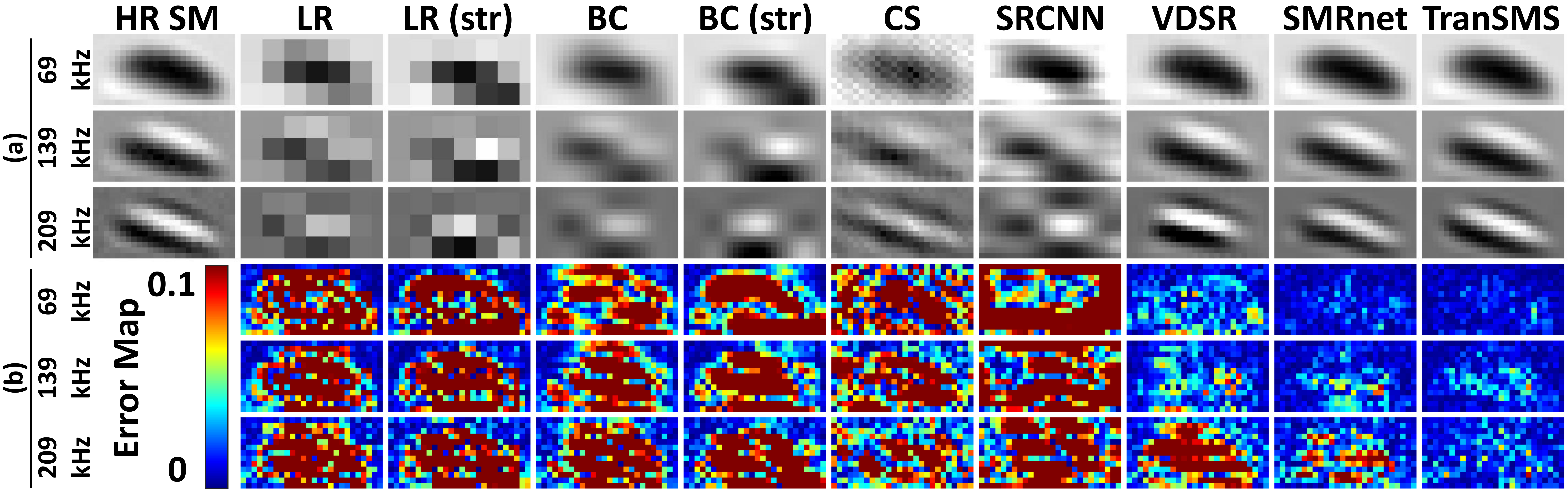}}
    \end{minipage}\hfill
\end{figure*}

\begin{figure*}[t]
\centering
    \begin{minipage}{0.4\textwidth}
\caption{Phantom images in the in-house dataset reconstructed with recovered HR SMs at $4\times$ SR and respective error maps are displayed, along with the benchmark reconstruction using the reference HR SM. \textcolor{black}{The length, inner radius and outer radius of the tubes were 20 mm, 2 mm and 4 mm, respectively.}
}
	\label{fig:image_reconAselsanFFL}
	\end{minipage}\hfill
	\begin{minipage}{0.6\textwidth}
	\centerline{\includegraphics[width=0.95\textwidth]{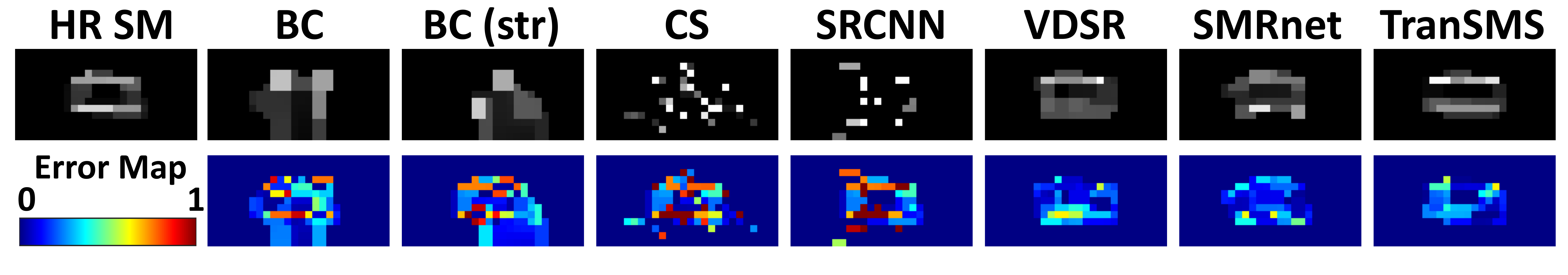}}
    \end{minipage}\hfill
\end{figure*}
\subsection{Analyses on In-house MPI Dataset}

\subsubsection{SF Variability}

Next, we demonstrated TranSMS on the in-house MPI dataset acquired on an FFL scanner. \textcolor{black}{We first assessed reliability against variability in SF gradients between SMs in the training versus test sets.} Table~\ref{table:comparison_AselsanNsy} lists nRMSE in SM recovery. TranSMS outperforms competing methods, with elevated benefits at higher SR factors. On average, it yields 3.99\% lower nRMSE than the second-best method. Recovered SMs are displayed in Fig.~\ref{fig:aselsanFFL8} at 4$\times$ SR. TranSMS yields the closest estimates to the reference HR SM, while 2d-SMRnet offers the second-best performance. Table~\ref{table:comparison_AselsanNsy} lists pSNR for image reconstruction using recovered HR SMs. TranSMS yields the highest pSNR,  outperforming the closest competitor by 1.59 dB pSNR on average. Reconstructions at 4$\times$ SR are shown in Fig.~\ref{fig:image_reconAselsanFFL} for the phantom in Fig. \ref{fig:inhouse_phantom}b. TranSMS produces a reconstruction with a high degree of similarity to the reference phantom image.

\begin{table}[b]
\centering
\caption{Average nRMSE ($\%$) in HR SM recovery / pSNR (dB) in image reconstruction for the in-house dataset at $2\times$-$8\times$ SR.} 
\label{table:comparison_AselsanNsy}
\begin{tabular}{|c|c|c|c|} 
\hline 
& $2\times$ & $4\times$ & $8\times$  \\ 
\hline 
Bicubic & 10.97\% / 16.25 & 36.26\% / 13.75 & 69.76\% / 4.25 \\
\hline 
Bicubic (str.) & 25.99\% / 15.84 & 68.40\% / 12.88 & 101.52\% / 5.63 \\
\hline 
CS & 15.31\% / 17.30 & 62.54\% / 7.68 & 104.00\% / 2.59 \\
\hline 
SRCNN & 43.79\% / 13.78 & 75.40\% / 6.46 & 437.66\% / 6.08 \\
\hline 
VDSR & 7.46\% / 20.24 & 14.78\% / 18.39 & 30.46\% / 10.21 \\
\hline 
2d-SMRnet & 6.65\% / 22.71 & 11.27\% / 19.38 & 56.06\% / 6.22 \\
\hline 
TranSMS & \textbf{6.03\% / 24.89} & \textbf{8.92\% / 19.68} & \textbf{21.47\% / 12.51} \\
\hline 
\end{tabular}
\end{table}

\subsubsection{Prospective Experiment}

\begin{figure}[t]
\centering
	\includegraphics[scale=0.23]{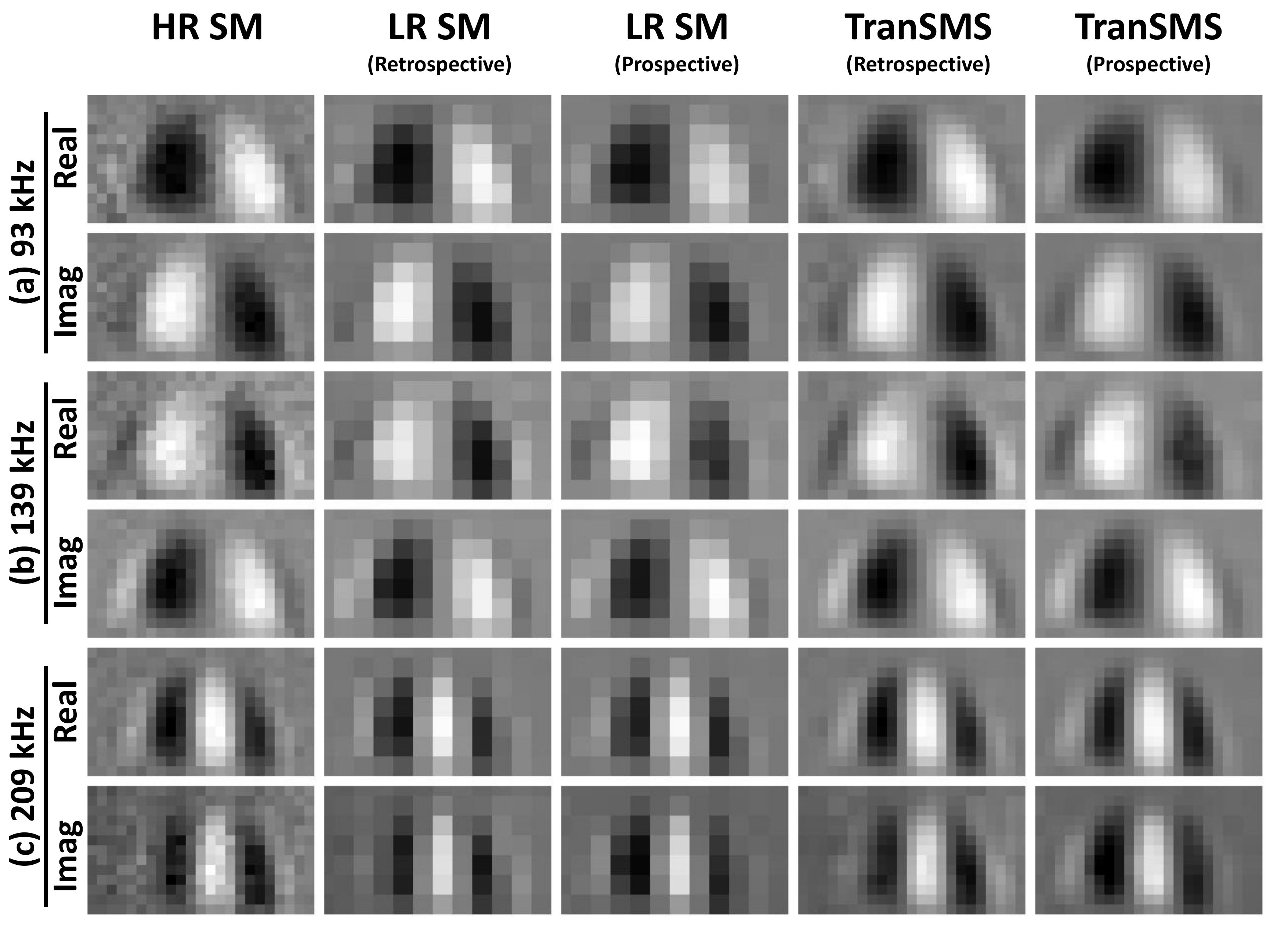}
	\caption{Prospective versus retrospective HR SM recovery in the in-house dataset for $2\times$ SR. \textcolor{black}{Real and imaginary parts of the SM rows corresponding to different harmonics (\textbf{(a) -- (c)}) at 0$^\circ$ are shown}.
	}
	\label{fig:myProspectiveResult}
\end{figure}

\begin{figure}[t]
\centering
	\includegraphics[scale=0.16]{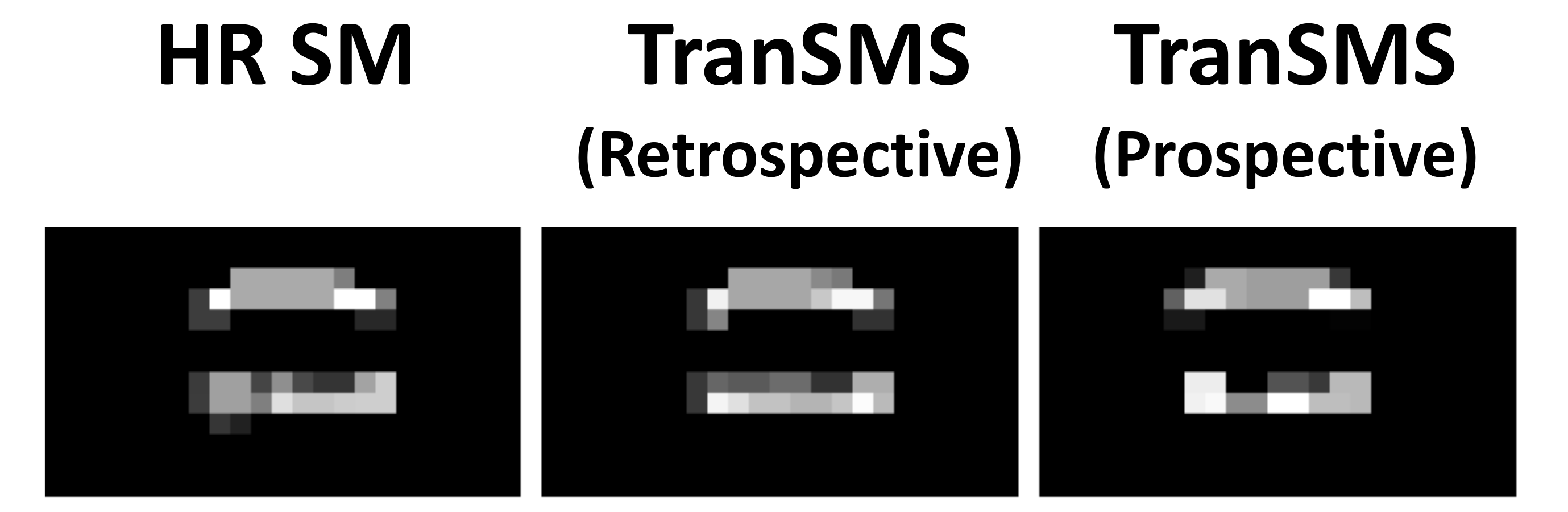}
	\caption{Phantom images for the in-house dataset based on prospectively versus retrospectively recovered HR SMs are shown along with the benchmark reconstruction using the reference HR SM. The upper tube is 23-mm long, the lower tube is 21-mm long with a 5-mm-long central stenosis region of 1-mm inner radius (1 pixel $:$ 2 mm).}
	\label{fig:myProspectiveResultPhantom}
\end{figure}

Finally, we demonstrated TranSMS on an LR SM prospectively measured using a large MNP sample spanning an LR voxel. The model trained on retrospectively downsampled LR SMs was used for inference. HR SMs recovered from retrospectively versus prospectively sampled LR SMs are displayed in Fig.~\ref{fig:myProspectiveResult} at 2$\times$ SR. As expected, HR SM based on prospectively sampling shows lower noise levels. Reconstructions using recovered HR SMs are shown in Fig.~\ref{fig:myProspectiveResultPhantom}. Reconstructions based on prospective sampling better preserve the stenosis in the phantom compared to retrospective sampling. \textcolor{black}{A quantitative metric cannot be reported for prospective sampling due to absence of ground truth. The theoretical improvement in measurement SNR with prospective sampling is 6dB at 2$\times$ SR. Based on simulation analyses, a crude expectation is that this improvement will translate onto 1-2dB difference in reconstruction performance (results not shown).}

\section{Discussion}
To accelerate MPI calibration, here we proposed to recover HR SM via a vision transformer model given LR SM measured with large MNP samples. Clear benefits were observed over strided or randomly undersampled measurements with small MNP samples \cite{superResolutionMPISM, omer_2015, Gungor_iwmpi}. In the simulated dataset, LR SMs were obtained by prospective downsampling of HR SMs by a factor of $S$ in each dimension. Signal power increases $S^4$-fold while the noise power remains constant, improving SNR by $10 \log_{10}(S^4)$, e.g., $\sim$24 dB at $4\times$ SR. \textcolor{black}{In experimental datasets, we trained TranSMS on LR SMs obtained by retrospective downsampling of HR SMs to mitigate potential biases due to nuisance variability between SM measurements. In that case, noise power increases $S^2$-fold, limiting the SNR improvement to $10 \log_{10}(S^2)$, e.g., $\sim$12 dB at $4\times$ SR. In prospective demonstrations with an LR SM measured using a large MNP sample, further benefits were observed due to improved SNR. An imaging system with robust mechanical positioning might further permit prospectively sampled LR SMs to be used for model training.} \textcolor{black}{Alternatively, a hybrid SM calibration can be performed to measure the MNP response via a magnetic particle spectrometer (MPS) where sample position is emulated using an offset field \cite{von2017hybrid}, or physical movement of the sample can be avoided via homogeneous focus fields to move the FFR \cite{Halkola_SCU}. Such SM measurements might also be super-resolved with TranSMS for further increasing the achievable resolution in HR SM.}

Here, we observe that HR SM recovery performance is higher in simulated versus experimental data, likely due to several factors. First, retrospective sampling of LR SM limits SNR benefits on experimental data. While non-stationary system parameters between separate measurements including spatial misalignment render quantitative evaluations of prospective sampling challenging, our preliminary results suggest further improvements in this case. \textcolor{black}{Second, the training sets are relatively compact for experimental datasets due to practical limitations regarding SM measurement, whereas the simulated dataset compiles a diverse collection of SMs that helps boost performance of learning-based models. The ablation study on training set size suggests diminishing returns beyond a certain number of training samples, yet future work is warranted to find the precise number of SMs needed for near-optimal performance in experimental datasets.} Third, our simulations do not consider nonidealities such as relaxation that can lower task difficulty. \textcolor{black}{Other influential factors include SF gradient strength that controls the spatial rate of change in SM, and MNP diameter that alters sensitivity to magnetic field and broadening of the response from FFR. In this work, we find elevated SM recovery errors at relatively high frequency components that show high spatial frequency information. This observation implies that factors that increase spatial resolution would also elicit increased task difficulty. Thus, in theory, task difficulty should elevate with higher SF gradients and larger MNP diameters.}

\textcolor{black}{All methods naturally show performance degradation towards higher SR factors as the spatial information in input measurements is constrained. For each method, performance becomes notably poor (e.g., $>$50\% nRMSE, $<$10dB PSNR) beyond an SR factor, which inevitably depends on the characteristics of SM measurements and varies across datasets. Yet, TranSMS generally shows higher resilience than competing methods against high SR factors.} \textcolor{black}{CS theory guarantees signal recovery for a near-isometric sampling matrix, and the recovery error depends critically on the number of measurements and SNR level. Here, we observe that CS performs competitively at 2$\times$ SR (i.e., 4-fold acceleration), albeit poorly at 4$\times$-8$\times$ SR (i.e., 16-fold to 64-fold acceleration). Our results are aligned with a recent study that assumes similar measurement SNRs for 2D imaging \cite{FastCalibrationSilbey}. While performant recovery using CS has been reported for 3D imaging under higher SNRs \cite{Weber_2015,hosvd_calibration}, different imaging scenarios are considered here.}     
\textcolor{black}{In Open MPI, we observed that 2d-SMRnet shows relatively higher performance differences between the two analyses that examine SF and MNP variability versus temporal variability compared to other methods. This could reflect an interaction between the sampling scheme and the training set. 2d-SMRnet uses strided LR SMs, and other methods use box-car downsampled LR SMs. In Open MPI, the training and test sets that we used for temporal variability analysis were collected under matching system parameters. In contrast, the training set that we used for SF and MNP variability analysis were collected with half-pixel shifts in the sampling grid, differing from the data that we used for the test set. The half-pixel shift during SM acquisition generates spatially overlapping measurements in the latter case, which does not strictly obey the downsampling model that TranSMS assumes. In contrast, these overlaps are not a concern for the strided measurements used by 2d-SMRnet. These differences likely contribute to the relative performance boost for 2d-SMRnet.}

Prior CS methods for accelerated MPI calibration perform randomly undersampled HR SM measurements with a small MNP sample, and recover SM by enforcing sparsity in a transform domain \cite{Lampe_2012, Weber_2015}. In contrast, TranSMS performs LR SM measurements with a large MNP sample to improve measurement SNR, and a data-driven recovery is performed between LR and HR SM instead of relying on sparsity assumptions. \textcolor{black}{An alternative for accelerated calibration is based on SR. In \cite{2d-SMRnet}, HR SM measurements are uniformly subsampled in a strided fashion, and HR SM is then recovered via a CNN. In \cite{superResolutionMPISM}, an LR SM is supplemented via an analytically-simulated HR SM, and a nonlinear optimization problem is solved to directly reconstruct an MPI image. Instead, we perform LR SM measurements with large MNP samples to improve SNR efficiency, we rely solely on measured SMs for reliability against system imperfections, and we leverage a novel vision transformer for SM recovery.} 

\textcolor{black}{It is possible to adopt the sampling strategy of TranSMS in previous CS- and SR-based methods. Note, however, that CS relies on measurements with incoherent interference in the sparsifying-transform domain. Large MNP samples would result in more structured spatial sampling, elevating coherence and degrading recovery. On the other hand, learning-based SR methods such as 2d-SMRnet can benefit from improved SNR efficiency due to large MNP samples similar to TranSMS.}

Voxel-sized undiluted MNP samples help improve SNR during calibration. Instead, clinically admissible intravenous administration of MNPs result in notably lower concentration per voxel during imaging. Particle interactions may alter MPI signal characteristics for highly concentrated samples, eliciting a discrepancy between calibration and imaging experiments \cite{ConcentrationDependent}. Because TranSMS enables SNR benefits due to larger MNP sample spanning an LR voxel, it might mitigate this issue by allowing calibration with diluted MNP samples.

Several developments can improve performance of TranSMS. TranSMS can be trained to perform SR on multiplexed LR SM measurements with multiple MNP samples to further calibration efficiency \cite{FastCalibrationSilbey}. Instead of using a uniformly filled LR voxel, optimizing the spatial distribution within an LR voxel could also increase the level of spatial information captured during calibration, and thereby improve SM recovery. Furthermore, measuring HR SM with larger MNP samples can increase SNR of reference images used to train TranSMS. However, the measured HR SM would be blurred due to box-car downsampling, so the DC module would have to be modified and deconvolution would be required to resolve HR SM at the intended resolution.

\section{Conclusion}

In this work, we introduced a novel MPI calibration approach that captures an LR SM with up to 64-fold larger MNP sample that spans across an LR voxel for improving measurement SNR. A novel transformer model is then leveraged to sensitively recover HR SM. TranSMS achieves lower error in SM recovery and higher quality in image reconstruction compared to state-of-the-art CS and SR methods. Therefore, it is a promising candidate to improve practicality of MPI.

\section{Appendix}
\label{app:proxProj}
\textbf{Proof:} Let us first define $\epsilon_i = \sqrt{m} \sigma_i$,  ${\mathbf{c}}_i = \mathbf{a} - \Tilde{\mathbf{a}}_i$ and $\Tilde{\mathbf{b}}_i = \mathbf{b}_i - \mathbf{D} \Tilde{\mathbf{a}}_i$.
The associated Lagrangian is, 
\begin{align}
L({\mathbf{c}}_i, \lambda) &= \Vert {\mathbf{c}}_i \Vert_2^2 - \lambda\left(\epsilon_i^2 - \Vert \mathbf{D}{\mathbf{c}}_i - \Tilde{\mathbf{b}}_i \Vert_2^2\right).
\end{align}
Considering the Lagrangian function, we get two conditions. In the first case, if $\Vert \Tilde{\mathbf{b}}_i \Vert_2 \leq \epsilon_i$, then ${\mathbf{c}}_i = 0$ and $\hat{\mathbf{a}}_{i} = \Tilde{\mathbf{a}}_i$. In the second case, we get the optimality conditions as:
\begin{align}
0 &= {\mathbf{c}}_i + \lambda \mathbf{D}^T \left(\mathbf{D}{\mathbf{c}}_i - \Tilde{\mathbf{b}}_i\right), &\epsilon_i^2 = \Vert \mathbf{D}{\mathbf{c}}_i - \Tilde{\mathbf{b}}_i \Vert_2^2\label{eq:lag1} 
\end{align}
Now using Eq.~\eqref{eq:lag1}, 
\begin{align}
 \lambda\mathbf{D}^T\Tilde{\mathbf{b}}_i &= \left(\lambda \mathbf{D}^T \mathbf{D} + \mathbf{I}\right) {\mathbf{c}}_i, \\
 {\mathbf{c}}_i &= \lambda \left(\lambda \mathbf{D}^T \mathbf{D} + \mathbf{I}\right)^{-1} \mathbf{D}^T \Tilde{\mathbf{b}}_i.
\end{align}
Using the Woodbury matrix identity and $\mathbf{D}\mathbf{D}^T = \mathbf{I}$:
\begin{align}
{\mathbf{c}}_i &= \lambda / (\lambda + 1) \mathbf{D}^T  \Tilde{\mathbf{b}}_i.
\end{align}
Now, let us put ${\mathbf{c}}_i$ back into Eq.~\eqref{eq:lag1}:
\begin{align}
\epsilon_i^2 &= \Vert \lambda / (\lambda + 1) \mathbf{D}\mathbf{D}^T  \Tilde{\mathbf{b}}_i - \Tilde{\mathbf{b}}_i \Vert_2^2, \\
\lambda / (\lambda + 1) &= 1 - \epsilon / \Vert\Tilde{\mathbf{b}}_i \Vert_2
\end{align}
Finally, the result becomes:
\begin{align}
{\mathbf{c}}_i &= \left(1 - \epsilon_i / \Vert\Tilde{\mathbf{b}}_i \Vert_2\right) \mathbf{D}^T  \Tilde{\mathbf{b}}_i, \\
\hat{\mathbf{a}}_{i} &= \Tilde{\mathbf{a}}_i + \left(1 - \sqrt{m} \sigma_i / \Vert\Tilde{\mathbf{b}}_i \Vert_2\right) \mathbf{D}^T \Tilde{\mathbf{b}}_i.
\end{align}

\bibliographystyle{IEEEtran}
\bibliography{IEEEabrv,refs}

\end{document}